\documentclass[aps,pra,superscriptaddress,showpacs,showkeys]{revtex4}
\usepackage[dvips]{graphicx}
\usepackage{longtable}
\usepackage{bm}
\usepackage{color}
\DeclareBoldMathCommand{\bfmu}{\mu} 
\def\mum{\rm\mu m}                  
\def\muA{\rm\mu A}                  
\def\muK{\rm\mu K}                  
\def\ie{{\it i.e.,\/}}              
\def\eg{{\it e.g.,\/}}              
\def\via{{\it via\/}}               
\def\GP{Gross-Pitaevskii}           
\def\CP{Casimir-Polder}             

\begin{document}

\title{Nanowire atomchip traps for sub-micron atom-surface distances}

\author{ R. Salem }	
	\thanks{Corresponding author}
	\email{salemr@bgu.ac.il}
	\affiliation{Department of Physics, Ben-Gurion University, Be'er Sheva 84105, Israel}
\author{ Y. Japha }
	\affiliation{Department of Physics, Ben-Gurion University, Be'er Sheva 84105, Israel}
\author{ J. Chab\'e }
	\affiliation{Department of Physics, Ben-Gurion University, Be'er Sheva 84105, Israel}	
\author{ B. Hadad }
	\affiliation{The Weiss Family Laboratory at the Ilse Katz Institute for Nanoscale Science and Technology, Ben-Gurion University, Be'er Sheva 84105, Israel}
\author{ M. Keil }
	\affiliation{Department of Physics, Ben-Gurion University, Be'er Sheva 84105, Israel}
\author{ K. A. Milton }
	\affiliation{Homer L. Dodge Department of Physics and Astronomy, University of Oklahoma, Norman, Oklahoma 73019, U.S.A.}
\author{ R. Folman }
	\affiliation{Department of Physics, Ben-Gurion University, Be'er Sheva 84105, Israel}

\begin{abstract}
We present an analysis of magnetic traps for ultracold atoms based on current-carrying wires with sub-micron dimensions. We analyze the physical limitations of these conducting wires, as well as how such miniaturized magnetic traps are affected by the nearby surface due to tunneling to the surface, surface thermal noise, electron scattering within the wire, and the Casimir-Polder force. We show that wires with cross sections as small as a few tens of nanometers should enable robust operating conditions for coherent atom optics (\eg\ tunneling barriers for interferometry). In particular, trap sizes on the order of the deBroglie wavelength become accessible, based solely on static magnetic fields, thereby bringing the atomchip a step closer to fulfilling its promise of a compact device for complex and accurate quantum optics with ultracold atoms.
\end{abstract}

\date{\today}

\pacs{42.50.Ct, 37.10.Gh, 12.20.-m}
\keywords{atomchips, ultra-cold atoms, nanowire, nano fabrication, atom optics, magnetic trapping, fragmentation, decoherence, trap loss, Casimir-Polder, Gross-Pitaevskii}

\maketitle

\section{Introduction\label{sec:introduction}}

Trapping atoms in magnetic traps using atomchips~\cite{RonRev, ReichelRev, fortagh} allows ultracold atoms or a Bose-Einstein Condensate~(BEC) to be manipulated and interrogated very close to the atomchip surface. Though outstanding achievements have already been made for atom-surface distances of~$\rm1-100\,\mum$, \eg\ spatial interference~\cite{JoergIFM,Ketterle_int} as well as hyperfine state interferometry~\cite{ReichelClock}, it remains of paramount importance to understand what ultimately limits the atom-surface distance. Further decreasing the atom-surface distance should increase trap gradients sufficiently to construct tunneling barriers with widths on the order of the atomic deBroglie wavelength, enabling \eg\ atomchip interferometry based solely on static magnetic fields. Such high trap gradients may also allow more robust atom-light interactions such as probing without heating in the Lamb-Dicke regime. Furthermore, sub-micron distances are also important for technological advantages such as low power consumption and high-density arrays of traps.

At small atom-surface distances, interactions with the nearby surface become important. For example, spatial and temporal magnetic field fluctuations, due to electron scattering and Johnson noise respectively, limit the minimum atom-surface distance, as they cause potential corrugations, spin flips, and decoherence. There have been several experiments utilizing cold atoms to study these interactions~\cite{tubinghan_frag, Hinds_spinflip, orsay1, MagScanN, simonSc, emmert}, and many suggestions on how to overcome their damaging effects~\cite{Valery, OrsayFragSup, fermani, YoniPRB, TalAni, CNT}.

Also becoming prominent for small atom-surface distances is the Casimir-Polder~(CP) force~\cite{sukenik}; normally attractive, it reduces the magnetic barrier and allows atoms to tunnel to the surface, as already observed \cite{vuleticCP,cornell}. At very small distances the atoms may also serve as a sensitive probe for surface phenomena; for example, plasmons are expected to affect the atomic external and internal degrees of freedom and may also become observable~\cite{IntravaiaCP}.

From the above it is evident that achieving small atom-surface distances would not only be advantageous for atom optics, but would also contribute to the fundamental study of surface phenomena. Finally, let us note that there are numerous ideas for bringing atoms closer to the surface~\cite{Ovchinnikov, Schmiedmayer, Shevchenko, rosenblit1, Ricci, Bender, Gillen}, all of which are, however, based on interactions with fields other than static magnetic fields, the latter being the focus of this work. We consider wires operating at room temperature, fabricated using standard methods, in contrast to suspended molecular conductors~\cite{peano, fermani} and superconductors, that reduce the \CP\ force and noise originating in the surface~\cite{valery2, emmert, roux, mukai, cano, Haroche09} respectively.

The paper is organized as follows: in Sec.~\ref{sec:motivation} we show qualitatively that creating static potential barriers on the scale of atomic deBroglie wavelengths, and therefore suitable for controlling tunneling, require micron or sub-micron atom-surface distances. In Sec.~\ref{sec:properties} we present the physical properties of gold nanowires, including a theoretical analysis of their resistivity. In Sec.~\ref{sec:potential} we analyze the potentials expected from such nanowires, including the \CP\ force and potential corrugation effects. We show that improved fabrication methods can overcome earlier limitations due to trapping potential roughness~\cite{orsay1}, which at sufficiently small separations would otherwise cause the trapped atomic cloud to break into smaller clouds (fragmentation). In Sec.~\ref{sec:loss}, we estimate trap lifetimes limited by atom losses due to noise-induced spin flips, Majorana spin flips, and tunneling. In particular, we show how the spin-flip rate induced by Johnson noise is reduced naturally by using very small amounts of material in the nanowires. We also consider the issue of decoherence. In Sec.~\ref{sec:example} we discuss a simple trap configuration based on a Z-shaped gold wire. We show that such a nanowire structure can generate static magnetic potentials, smooth enough for trapping a BEC at sub-micron atom surface separations. Finally, in Sec.~\ref{sec:anisotropic} we briefly discuss nanowire traps fabricated by more exotic materials.

\section{Static magnetic potentials for atom interferometry\label{sec:motivation}}

Two technical characteristics of potentials that are required in order to study atom interferometry can be described in the following way: first, the potential barrier between adjacent wells should be sufficiently low or narrow so that the tunneling rate is comparable to, or faster than, typical experimental or dephasing time scales; and second, that this tunneling rate can be controlled with experimentally accessible currents and fields.

Largely because of the weak~$1/r$ dependence of the magnetic potential on the atom-wire distance, these tunneling conditions require very short distances. To quantify this, we construct a simple waveguide potential using a single atomchip wire (in the x-direction) and an external bias field; current through a second atomchip wire (in the y-direction) is added to generate a simple potential barrier in a right-angle~``X'' wire configuration~\cite{ReichelIFM}. This configuration incidentally is exactly opposite to the ``dimple'' configuration recently used for compressing atomchip traps~\cite{Horikoshi, Anderson-poster}.

\begin{figure}[ht]
	\includegraphics[width=0.45\textwidth]{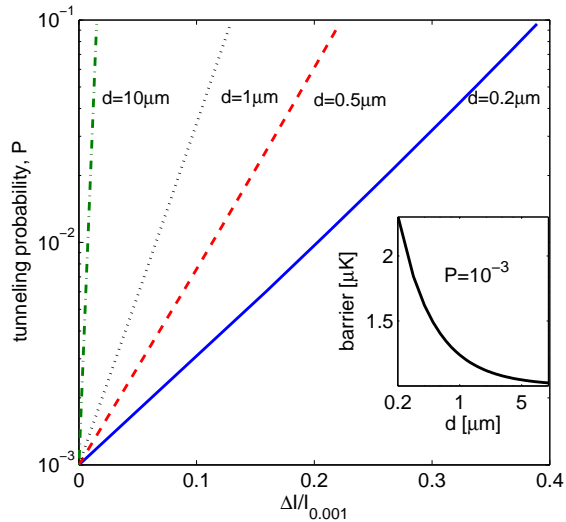}
		\caption{(Color online) Tunneling probability through a barrier at several heights~$d$ as a function of the change in the current~$\Delta I$ through the control wire, relative to the current~$I_{0.001}$ for a probability~$P(I)=0.001$. A kinetic energy of~$\rm1\,\muK$ is assumed for a single atom of~$\rm^{87}Rb$. For this X-wire configuration, changing the current by a few percent causes a drastic change in the tunneling probability for~$d=\rm10\,\mum$. Good control over the tunneling probability requires the height to be about~${d\lesssim\rm1-2\,\mum}$. The inset shows the potential barrier required to maintain a tunneling probability of~$0.001$ as a function of the atom-surface separation~$d$: for smaller~$d$, a higher barrier is required (as the barrier becomes thinner) so better tunneling control is attained.  The motivation for small atom-surface distances is quantified further in Sec.~\ref{subsec:engineering}.}	
		\label{fig:barrier}
\end{figure}

The magnetic potential in the~$x-$direction, generated by the crossing wire, is given by:
\begin{equation}
	V(x)=\mu_A B_0+\frac{\mu_A \mu_0 I}{2\pi}\ \frac{z}{z^2+x^2},
	\label{eq:barrier}
\end{equation}
where~$\mu_A$ is the atomic magnetic dipole moment along the direction of the Ioffe field~$B_0$, $I$ is the current in the crossing wire, $\mu_0$ is the permeability of free space, and~$z$ is the distance of the atom from the atomchip surface. A one-dimensional single-particle tunneling probability through the barrier can then be calculated in the~WKB approximation as
\begin{equation}
  P=\exp\left(-\frac{1}{\hbar}\int_{-x_E}^{x_E}dx \sqrt{2m[V(x)-E]}\right),
 	\label{eq:Tunnel}
\end{equation}
where~$E$ is the kinetic energy of the atom and~$V(\pm x_E)=E$. Assuming a kinetic energy of~$E=1\,\muK$ for a~$\rm^{87}Rb$ atom (corresponding to a free-particle deBroglie wavelength of~${\approx\rm0.33\,\mum}$) in the~$\left|F=2,m_F=2\right\rangle$ state, we may then easily calculate the change required in the current~$I$ that causes a given proportional change in the tunneling probability, as a function of the atom-surface distance~$z$. The results of this calculation are shown in Fig.~\ref{fig:barrier} for changing the tunneling probability from~0.001 to some higher probability. The calculation suggests that control over the tunneling probability requires a distance~$z$ on the order of~$\rm1-2\,\mum$ for experimentally reasonable values of current control. In the simple model of Eq.~(\ref{eq:barrier}), this corresponds to a barrier half-width of~$\rm2-4\,\mum$, comparable to experiments that have observed interference between adjacent wells with the addition of non-static fields~\cite{JoergIFM,Ketterle_int}. Thus, the desired static magnetic potentials can be generated only if atoms can be brought down to micron or sub-micron distances above the wires on the atomchip surface, at which point the tunneling rate can be tuned over an experimentally useful dynamic range by adjusting the current in the crossing wire. One may then envisage interferometric devices such as the ones we have proposed in Refs.~\cite{DanielPRL} and~\cite{YoniPRL}.

It is well known that, to avoid finite size effects which degrade the trap gradient, the wire size should be on the same scale as the atom-surface distance, \ie\ for the above noted heights of $d\lesssim\rm1-2\,\mum$ (see Fig.~\ref{fig:barrier}) one requires a micron-scale wire. As will be shown in the following, it is advantageous to utilize even smaller wire dimensions, namely nanowires. This will enable improving operational parameters at the above heights, or decreasing the atom-surface distance even further without hindering effects.

\section{Physical properties of thin wires\label{sec:properties}}

\subsection{Wire fabrication and characterization\label{subsec:fab}}

\begin{figure}[ht]
 	\includegraphics[width=0.85\textwidth]{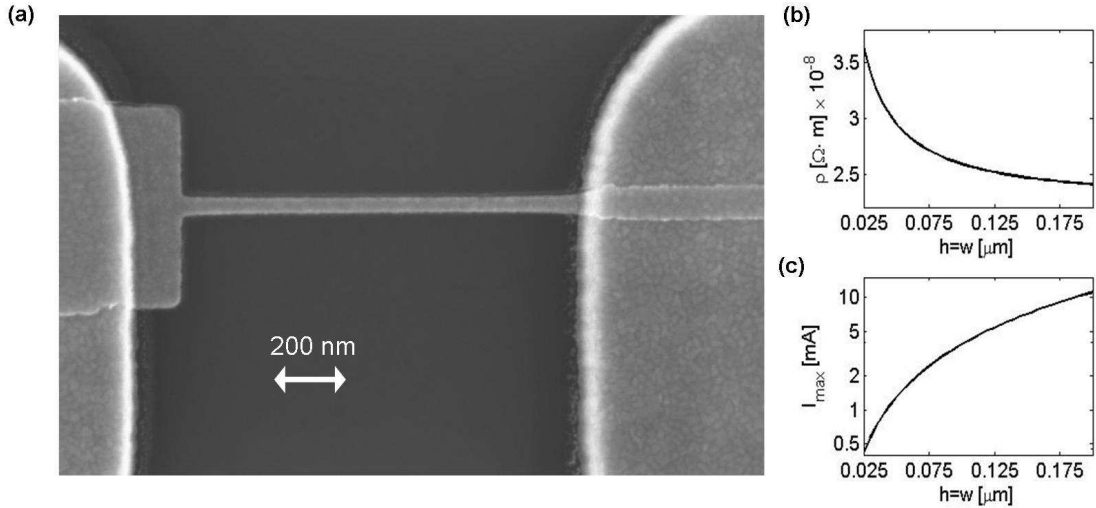}
	\caption{(a)~Scanning electron microscope~(SEM) image of a~$\rm2\,\mum$ long, $\rm20\,nm$ thick and~$\rm50\,nm$ wide gold wire. Unless otherwise noted, the wires considered in this paper have square cross sections. (b)~Calculated dependence of resistivity on wire dimensions, based on the Fuchs-Sondheimer surface scattering model~\cite{Fuchs, Sondheimer} of Eq.~(\ref{eq:sondheimer}). (c)~Maximum current considered safe for atomchip wire operation, calculated for different wire cross-sections from Eq.~(\ref{eq:wire heating}), assuming the nanowire resistivity~$\rho$ shown in~(b) and the temperature coefficient~$\alpha$ for bulk gold.
\label{fig:rho}}
\end{figure}

In order to study the possibility of trapping atoms using nanowires, we first discuss the fabrication feasibility. An example of one such (short) wire,~$\rm20\,nm$ thick and~$\rm50\,nm$ wide, is shown in Fig.~\ref{fig:rho}(a).
This wire was prepared by us in a relatively simple two-step process involving optical lithography (for external connection) followed by electron-beam lithography (for the nanowires). A~Si substrate with a well-defined oxide layer of~$\rm100\,nm$ thickness and a thin~$\rm5\,nm$-thick Ti adhesion layer is spin-coated with image reversal photoresist, which is then exposed to ultraviolet light through a mask. After developing, a~$\rm5\,nm$-thick Ti seed layer, followed by a~$\rm200\,nm$-thick Au layer, is then evaporated onto the sample and the undeveloped areas lifted off, leaving large areas for connection to external testing equipment. The sample is then spin-coated with a layer of~PMMA, and the nanowire plus several of the interconnects are patterned by electron-beam lithography. After developing, gold is evaporated onto the sample with the desired thickness and the fabrication is completed with a final lift-off process. This fabrication process can easily be integrated with any atomchip design.

Using~SEM images of the fabricated wires, we measured the edge roughness of the resulting wires, as shown in Fig.~\ref{fig:rho}(a). The spectrum of this edge roughness can be characterized as frequency-independent (``white noise'') with a measured root-mean-square deviation of~$\rm2\,nm$ for wavelengths of~$\rm100-800\,nm$.

\subsection{Wire resistance calculations}\label{subsec:resistance}

The resistivity of a nanowire increases beyond the bulk resistivity as the cross-section dimensions become comparable to the mean free path~$l$ of an electron ($l\approx\rm40\,nm$ for gold at room temperature  \cite{Ashcroft}). In such a small wire the resistivity may increase significantly~\cite{DurcanWelland, Engelhardt}. To estimate the change of resistivity in a nanowire, we follow the theoretical model of Fuchs and Sondheimer~\cite{Fuchs,Sondheimer}, which was extended by Chambers~\cite{Chambers}. This model is supported experimentally for gold nanowires~\cite{DurcanWelland}. For wire dimensions on the order of the grain size, a supplementary model by Mayadas and Shatzkes~\cite{Mayadas} is needed in order to account for scattering at grain boundaries. For the simple fabrication process we have described, the measured grain size is about~$\rm20\,nm$, so for wire dimensions above this size we can attribute the increase in nanowire resistivity solely to scattering at the walls as in the Fuchs-Sondheimer model~\cite{Durkan2}. 

For atomchip experiments, we are interested in the current density in the wire and not only the wire resistance. Therefore we give the current density in a wire (along~$\hat x$) of width~$w$ (along~$\hat y$) and thickness~$h$ (along~$\hat z$) as
\begin{equation}
J(y,z) = J_0\left[1 - s(y;z)-s(w-y;z)-s(z;y)-s(h-z;y)\right],
\end{equation}
where $J_0=I/wh$ is the current density expected in the absence of surface scattering, and~\footnote{Note the changes we have made to Eq.~(1) in~\cite{DurcanWelland} and to Eq.~(2) in~\cite{Engelhardt}.}
\begin{equation}\label{eq:sondheimer}
s(y;z)=\frac{3}{4\pi}
\int\limits_{-\arctan\left(z/y \right)}^{\arctan{\left[(h-z)/y\right]}}
{d\phi\ \int_0^{\pi} d\theta\ \sin\theta \cos^2\theta \
\exp\left(\frac{-y}{l\sin\theta\cos\phi}\right)}
\end{equation}
corresponds to scattering at the~$y=0$ boundary, $s(w-y;z)$ corresponds to scattering at the~$y=w$ boundary, and~$s(z;y)$ and~$s(h-z;y)$, corresponding to scattering at the~$z=0$ and~$z=h$ boundaries respectively, are obtained by replacing~$y$ by~$z$ and~$h$ by~$w$. The resulting resistivity is given by
$\rho/\rho_0=J_0/\int\int dy\ dz\ J(y,z)$, where~$\rho_0$ is the bulk resistivity. It follows that the current density at the metallic layer near the boundary drops to~$\frac{1}{2}$ its value far from the boundary. To account for surface scattering, one assumes a fraction~$p$ ($0\leq p\leq 1$) of specular reflection events at the boundaries; then the value of the resistivity is given by a series expansion
\begin{equation} \left(\frac{\rho_0}{\rho}\right)_{p,l}=(1-p)^2\sum_{k=1}^{\infty} kp^{k-1}
\left(\frac{\rho_0}{\rho}\right)_{p=0,l/k},
\end{equation}
where~$(\rho_0/\rho)_{p=0,l/k}$ is the resistivity calculated for a wire with totally diffusive scattering at the boundaries~($p=0$) and a mean free path~$l/k$. Measurements of resistivity of thin gold wires are well reproduced by a theory assuming~$p=\frac{1}{2}$~\cite{DurcanWelland, Engelhardt}. Figure~\ref{fig:rho}(b) shows that the calculated resistivities for wires with square cross-sections increase by up to about~50\% for cross-sections down to~$\rm25\,\mum$.

\subsection{Current limitations}\label{subsec:limitations}

Forming magnetic traps deep enough to hold ultracold atoms near the surface of an atomchip requires
sufficiently large currents in the microfabricated wires to ensure that the trapping potential overcomes the gravitational force, the \CP\ attraction to the chip surface, and the kinetic and repulsive energy of the atoms.
However, if the current is too high, the wire overheats and may eventually break down~\cite{Sonke}. The wire temperature is determined by the balance between ohmic heating (whose power dissipation per unit area is~$h\rho J^2$) and the rate of heat conduction to the wafer per unit area~$-\kappa\Delta T$, where~$\kappa$ is the thermal contact resistance of the wire-wafer interface and~$\Delta T=T-T_0$ is the difference between the temperature~$T$ of the wire and the temperature~$T_0$ of the wafer (typically room temperature). The heat capacity of nanowires is so small that the wire reaches its maximum temperature very rapidly; approximating the temperature dependence of the resistivity as that of bulk gold, whose linear coefficient is~$\alpha=\rm0.0037\,K^{-1}$, we obtain the current density required to heat a given wire by a temperature~$\Delta T$ as~\cite{Sonke}
\begin{equation}
	J_{\rm max}=\sqrt{\frac{\kappa\Delta T_{max}}{h\rho(T_0)[1+\alpha\Delta T_{max}]}}~,
	\label{eq:wire heating}
\end{equation}
thus showing that thin wires allow higher current densities. On the other hand, if the wire cross-section is on the order of the mean free path of the electrons, the rise in the resistivity due to surface scattering [Fig.~\ref{fig:rho}(b)] may limit this advantage. In Fig.~\ref{fig:rho}(c) we present the calculated maximum current density for different wire cross-sections using an estimated value for~$\kappa=4\times10^6\rm\,Wm^{-2}K^{-1}$ from Ref.~\cite{Sonke}, and assuming that we limit the rise in resistivity (due to heating) to~50\%, which we consider to be within safe operating limits for thin atomchip wires~\cite{Sonke}.  This limitation in the resistivity change corresponds to heating by $\Delta T=1/2\alpha=135^{\circ}$.

When considering a specific example (Sec.~\ref{sec:example}), we will show that currents sufficient for generating atomchip traps may be an order of magnitude lower than the limits shown in Fig.~\ref{fig:rho}(c).

\section{Atomic trapping potential\label{sec:potential}}

In this section we describe two prominent effects influencing the static potential at sub-micron atom-surface distances generated by nano-scale wires, namely corrugations due to electron scattering, and tunneling to the surface or the nanowire, through the magnetic potential, due to the \CP\ force.

\subsection{Potential Corrugations\label{subsec:corrugations}}

One of the limiting factors when trapping or guiding atoms in a magnetic potential generated from a current carrying wire is the static potential corrugation due to current deviations~\cite{simonSc,YoniPRB}. Such current irregularities are produced by wire imperfections, namely, geometrical properties (wire edge roughness and surface roughness), and internal bulk inhomogeneities. Since atomchip traps are formed by canceling the magnetic field~$B_y$ generated by the current density~$J^0_x$ at a specific distance from the wire~$d$, the minimum of the trapping potential lies along the wire direction~$\hat{x}$. Variations in this potential~$\delta B_x(x)$ are then directly related to changes in the direction of the magnetic field generated by the wire imperfections.

In previous work~\cite{simonSc,YoniPRB}, we concluded that internal bulk inhomogeneities play a minor role in thin wires~($h<\rm250\,nm$). For wide wires, surface roughness dominates the potential corrugation, but as we show below, edge roughness dominates for narrow wires. Consequently, in this work we need to consider only current deviations due to edge roughness since all of the nanowires considered are thinner and narrower than~$h\approx w<\rm250\,nm$.

Let us consider a fabricated metal wire carrying a total current~$I$. It extends along the~$\hat{x}$ direction and has a width~$w$ along~$\hat{y}$ and thickness~$h$ along~$\hat{z}$. The boundaries of the wire are located at
$y=\pm w/2+\delta y_{\pm}(x,z)$ and $z=\pm h/2+\delta z_{\pm}(x,y)$. The corrugations of the wire boundaries~$\delta y_{\pm}$ and~$\delta z_{\pm}$ can be expanded as
\begin{equation}
  \delta y_{\pm}(x,z)=\sum_{n=-\infty}^{\infty}e^{2\pi i n z/h}\sum_k e^{ikx} \delta y^{\pm}_n(k)~~{\rm and}~~
  \delta z_{\pm}(x,y)=\sum_{m=-\infty}^{\infty}e^{2\pi i m y/w}\sum_k e^{ikx} \delta z^{\pm}_m(k).
\end{equation}
A linear theory for small corrugations predicts that the effect of each spectral component of the corrugation is responsible for a corrugation of the magnetic field near the atomic trap center with a similar wavelength~$2\pi/k$ along the~$x$ direction. However, the effect of components with wavelength much shorter than the distance~$d$ between the wire and the atomic trap (on the order of hundreds of nanometers or more) drops exponentially as~$e^{-|k|d}$ so that here we will only be interested in corrugations whose wavelengths are a few hundred nanometers or longer. We may then neglect the effect of spectral components on the order of the wire width or thickness and consider only corrugation terms with~$m=0$ and~$n=0$, \ie\ we may assume that~$\delta y_{\pm}$ and~$\delta z_{\pm}$ depend only on~$x$.

Corrugations of the magnetic field along the main trapping axis~$x$ above the center of a wire with geometrical perturbations are given by the Biot-Savart law as
\begin{equation}
  \delta B_x({\bf r})=\frac{\mu_0}{4\pi}\int d^3{\bf r}'\left[\delta J_y({\bf r}')
  \frac{\partial}{\partial z'}-\delta J_z({\bf r}')\frac{\partial}{\partial y'}\right]\frac{1}{
  {| \bf r- \bf r}'|}~,
  \label{eq:BS}
\end{equation}
where~$\delta J_y,~\delta J_z$ are the transverse current fluctuations in the wire. At the point exactly above the center of a nominally symmetric wire, it follows that only the symmetric components of~$\delta J_y$ 
[$\delta J_y(y)=\delta J_y(-y)$] and the anti-symmetric components of~$\delta J_z$ 
[$\delta J_z(y)=-\delta J_z(-y)$] contribute to the magnetic field. The fabrication process typically provides wires whose edge corrugations are much larger than their top or bottom surface corrugations, so that the symmetric part of~$\delta J_y$ is the major contribution to the magnetic field fluctuations.

Ohmic theory, which is adequate when the width and thickness of the wire are much larger than the electron mean free path and whose use we justify below, predicts that for wavelengths longer than the wire width or thickness the symmetric~$y$-current fluctuations in the wire have the form
\begin{equation}
  \delta J^{\rm sym}_y(x,y)=iJ^0_x \sum_{k\neq 0} k\ e^{ikx}\frac{(\delta y^+_k+\delta y^-_k)
  e^{-|k|w/2}}{1+e^{-|k|w}}\cosh(ky),
\end{equation}
such that in the limit where~$|k|w\ll 1$, $\delta J_y(x,y)\sim J_0 \partial y_{\rm center}/\partial x$, where $\delta y_{\rm center}=(\delta y_+ + \delta y_-)/2$ is the position of the actual center of the wire at a given point~$x$. Substituting this limit into Eq.~(\ref{eq:BS}) while assuming small deviations of the wire edges from their nominal position, and assuming that~$w\ll z$, \ie\ the width of the wire is much smaller than the distance of the atom to the wire, we obtain the following expression for the magnetic field corrugations above the wire
\begin{equation}
  \delta B_x(x,0,z)=\frac{iI\mu_0}{2\pi} \sum_k e^{ikx} \ k^2\ \delta y_{\rm center}(k) \ {\rm K}_1(|k|z),
  \label{eq:dBx}
\end{equation}
where $I=\int dy\int dz \ J(y,z)$ is the total current in the wire and~${\rm K}_1(kz)$ is the modified Bessel function, which may be approximated by ${\rm K}_1(u)\approx (e^{-u}/u)\sqrt{1+\pi u/2}$. Our model for the fluctuation spectrum assumes that $\delta y_c(k)=\delta y_0(k_0/k)^{\alpha}e^{i\varphi}$, where~$\delta y_0$ is the edge fluctuation at some wavevector~$k_0$ and then~$\delta y_c^{rms}$ can be obtained by summing this spectrum over all~$k$. Typically,~$\alpha$ is a number between~0 (``white-noise spectrum'') and~1 (``1/f spectrum''), while~$\varphi$ is a random phase. It follows that the root-mean-square value of the field fluctuations is given by
\begin{equation}
\langle\delta B_x^2\rangle=\left(\frac{I\mu_0\delta y_0 k_0^{\alpha}}{2\pi}\right)^2
\sum_k k^{4-2\alpha} {\bf K}_1(k|z|)^2.
\end{equation}
If we assume that the distance~$|z|$ is much shorter than the length~$L$ of the measured wire,
we obtain
\begin{equation}
 \frac{\delta B_x^{rms}}{B_0}\approx
A(\alpha)\frac{\delta y_c^{rms}}{(2z)^{3/2-\alpha}},
\label{eq:fragrms}
\end{equation}
where $B_0=I\mu_0/2\pi z$ is the regular magnetic field at the $y$ direction.
Here $A(\alpha)$ has units of (length)$^{1/2-\alpha}$ and is given by
\begin{equation}
A(\alpha)^2=\frac{L/\pi}{\sum_k k^{-2\alpha}}
\left[1+\frac{\pi}{4}(3-2\alpha)\right]\Gamma(3-2\alpha),
\end{equation}
where the sum is over $k$ values taking integer multiples of~$2\pi/L$ up to a cutoff~$k_{max}=2\pi/\lambda_{min}$. Typical values of this sum for~$\alpha=0$ and~$\alpha=1$ are~$\sum_k=L/\lambda_{min}$ and~$\sum_k k^{-2}=L^2/24$ respectively, when~$k_{max}\rightarrow\infty$.

The same result should be obtained if we consider diffusive surface scattering. As we have seen in Sec.~\ref{subsec:resistance}, in nano-sized wires, the conductivity near the boundary is reduced by diffusive surface scattering (with a typical exponential decay length $l$ from the wire edge). This means that diffusive scattering is limited to a region of dimension smaller than $l$. At the same time, the corrugation wavelengths~$2\pi/k$ relevant at the atom position, \eg\ similar to or larger than the atom-surface distance, induce current density directional daviations away from the edge with an exponential decay length of~$1/k$. As~$1/k \gg l$ most of the current will follow the corrugations of the boundary even in the case of surface diffusive scattering, such that the resulting $y$-current fluctuations will again generate transverse components of the current proportional to the derivative~$\partial\delta y_{\rm center}/\partial x$. We thus use the ohmic theory whose general form was developed in Ref.~\cite{YoniPRB} to calculate the magnetic field corrugations above the wire.

\begin{figure}[ht]
		\includegraphics[width=0.45\textwidth]{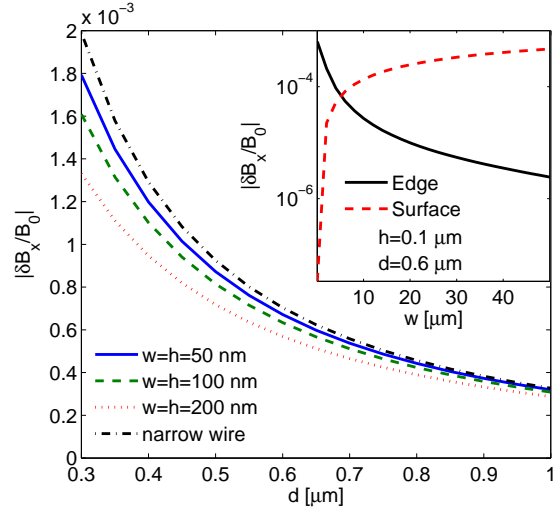}
	\caption{(Color online) Directional variation of the magnetic fields $|\delta B_x / B_0|$, calculated from Eq.~(\ref{eq:BS}) as a function of the atom-surface distance~$d$. We consider wires with square cross-sections of~$\rm50-200\,nm$ and the narrow-wire approximation presented in Eq.~(\ref{eq:fragrms}). The same edge roughness is used for calculating the magnetic variations for all the wires. The small differences amongst the wires, despite relatively higher edge roughness of the narrower wires, corresponds to Eq.~(\ref{eq:fragrms}) in which only the absolute quantity~$\delta y_c^{rms}$ appears. These differences are smallest for~$d\gg w$ and become larger as~$d$ approaches~$w$. The inset shows the directional variation of the magnetic field at a fixed height of~$d=\rm0.6\,\mum$ and for a fixed wire thickness of~$h=\rm0.1\,\mum$, where we plot the influence of edge roughness (solid curve) and surface roughness (dashed curve) over a wide range of wire widths~$w$. The effect of the surface roughness drops strongly for narrower wires, since long wavelengths of the magnetic corrugations are suppressed~\cite{YoniPRB}. For the nanowires considered herein, magnetic variations are completely dominated by edge roughness.}
  \label{fig:fragmentation}
\end{figure}

 In Fig.~\ref{fig:fragmentation} we present calculated directional variations of the magnetic field~$|\delta B_x/B_0|$, generated by the trapping wire, as a function of the height~$d$ for several wire cross-sections. The edge roughness amplitude is measured from our fabricated wires and was found to be frequency-independent [$\alpha=0$ in Eq.~(\ref{eq:fragrms})] with a measured root-mean-square deviation of~$\rm2\,nm$ between~$\rm100-800\,nm$~\footnote{We note that in the case of edge roughness with~$1\over f$ power spectrum ($\alpha=1$), the directional variations of the magnetic field~$\delta B_x/B_0$ will be an order of magnitude higher~($8\times10^{-3}$ compared to~$7\times10^{-4}$ at~$d=\rm0.6\,\mum$), and will lead to significantly larger density perturbations.}. In accordance with Eq.~(\ref{eq:fragrms}), we see that, for a given edge roughness~$\delta y_c^{rms}$, smaller wires  produce only slightly larger magnetic corrugations. The effect of such magnetic corrugations on the atomic density will be discussed in Sec.~\ref{sec:example}. We also see that the influence of the surface roughness~$\delta z_{\pm}(x,y)$ is negligible for the narrow wires discussed in this work due to the suppression of long wavelengths in the magnetic corrugations~\cite{YoniPRB}.

\subsection{Engineering longitudinal potential variations by nanowire shaping\label{subsec:engineering}}	
Shaping the nanowire edges may be used for creating potential variations desired for manipulating atoms near the atomchip surface. Having characterized the dependence of magnetic field variations on wire edge imperfections, we may now discuss quantitatively the deliberate ``tailoring'' of magnetic trapping potentials by engineering wire edge profiles. For the purposes of this study, as noted in the Introduction, we are particularly interested in potentials with sufficient variation so that tunneling barriers can be controlled. This is the main advantage of trapping atoms close to the trapping wire. Supplementing the motivation for a small atom-surface distance presented in Sec.~\ref{sec:motivation}, we now wish to determine the highest ``potential resolution'', \ie\ the smallest distinguishable distance between adjacent wells separated by static tunneling barriers, as a function of~$d$.

As a test case for quantifying this potential resolution, we consider a configuration in which a thin wire is curved with a certain periodicity~$\lambda$ that corresponds to a wave-vector $k=2\pi/\lambda$. If the amplitude of this curvature is small with respect to the wavelength, then the foregoing discussion implies that the magnetic field above the wire is given by a single~$|k|$ component in Eq.~(\ref{eq:dBx}), and then~$V(x)=V_0\cos kx$, where $V_0=\mu_A\mu_0 Ik^2\delta y_{\rm center}{\rm K}_1(kz)$.
\begin{figure}[ht]
	\includegraphics[width=0.45\textwidth]{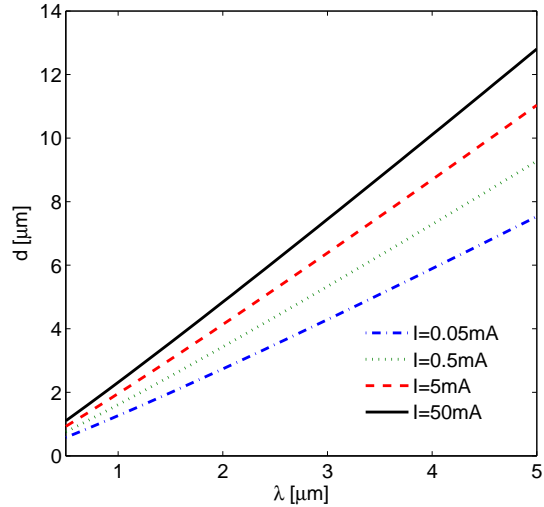}
\caption{(Color online) Potential spatial resolution achievable with wire currents from~$\rm0.05-50\,mA$. We present the maximum atom-surface distance~$d$ for which the longitudinal barrier between two adjacent minima in a periodic potential is at least twice the energy of the longitudinal ground state. Obtaining static magnetic potential features with a resolution on the order of the deBroglie wavelength, \ie\ for a potential periodicity on the order of~$\lambda\approx\rm1\,\mum$, requires the atom-surface distance to be~$d\lesssim\rm2\,\mum$. Wire currents of~$0.5$ and~$\rm5\,mA$ are the maximum currents that can be sustained through~$20$ and~$\rm100\,nm$ atomchip wires, respectively [Fig.~\ref{fig:rho}(c)]. The~$\rm0.05\,mA$ curve is useful when discussing a specific example of an atomchip trap (Sec.~\ref{sec:example}). Increasing the current by three orders of magnitude to~$\rm50\,mA$ serves to increase the required height by just a factor of two, despite being well beyond a safe atomchip current even for a~$\rm200\,nm$ nanowire.}
  \label{fig:PotRes}
\end{figure} 

At the minima of such a periodic potential, the longitudinal frequency is~$\omega=\sqrt{V_0 k^2/m}$, where~$m$ is the atomic mass. In order to engineer potential barriers between adjacent minima higher by a factor of say,~$\eta$ than the single-atom ground state energy, we require~$2V_0>\frac{1}{2}\eta\hbar\omega$, or~$V_0>(\eta^2/16)\hbar^2k^2/m$. In Fig.~\ref{fig:PotRes} we show the maximum atom-surface distance for which a longitudinal barrier with~$\eta=2$ can be obtained. These curves show that the maximum atom-surface distance still allowing tunneling control is on the order of the potential periodicity~$\lambda$. Designing the edges of a wire as the sum of different modulations therefore allows engineering of any periodic potential up to a resolution determined by the atom-surface distance. Consequently, as also seen in Sec.~\ref{sec:motivation}, atom-surface distances of~$\rm1-2\,\mum$ (or sub-micron distances in some cases) will be required to fully exploit the potential of an atomchip based on static magnetic fields.

\subsection{Attractive Casimir-Polder potential\label{subsec:CP}}

The \CP\ potential between a polarizable atom and dielectric or conducting objects~\cite{casimir} is one of the fundamental outcomes of zero-point vacuum fluctuations. It emerges from the fact that a dielectric or conducting object modifies the modes of the electromagnetic~(EM) field in its vicinity, modes which interact with the atomic polarization. In our case, an attractive \CP\ potential arises from the conducting gold nanowire and from the~Si wafer coated with a~$\rm100\,nm$-thick SiO$_2$ layer (used to prevent electrical shorts). The \CP\ potential reduces the potential barrier for tunneling to the surface, thereby limiting the possibility of trapping atoms near the surface~\footnote{Note that numerous ideas on how to alter the \CP\ force exist~\cite{IntravaiaCP, LeonhardtCP, MundayCP, YannopapasCP}, and may, if proven successful, enable decreasing the atom-surface distance even further.}.

The~EM modes of the combined surface+wire system are not analytically solvable and we will therefore carry out a separate examination of the \CP\ potential emerging from the~Si+SiO$_2$ planar wafer, as discussed in earlier
work~\cite{CNT}, and from a simplified model that takes the wire as a perfectly conducting circular cylinder of a certain diameter. We then take the sum of the two contributions as an estimate for the combined potential as a sort of pairwise additive approximation, PAA. Based on our earlier experience from the planar two-layer system, we anticipate that this approach should at least give the right order of magnitude for the accurate \CP\ potential.  

In general, the \CP\ potential may be written in the form
\begin{equation}
U_{CP}({\bf r})=i\hbar\int_{-\infty}^{\infty} d\omega\ \alpha(\omega)\left[\Gamma({\bf r},{\bf r},\omega)
-\Gamma_0({\bf r},{\bf r},\omega)\right],
\label{cp_general}
\end{equation}
where~$\alpha(\omega)$ is the frequency-dependent atomic polarizability and~$\Gamma({\bf r},{\bf r},\omega)$ is the trace over the Green's tensor of the electromagnetic field at the same point~${\bf r}$, with~$\Gamma_0$ being the Green's tensor in empty space, responsible for a space-independent Lamb shift. For distances from the dielectric or conducting object much larger than~$\lambda_0/2\pi$, where~$\lambda_0$ is the wavelength corresponding to the lowest optical transition frequency, the \CP\ potential generated by a planar structure made from a layer of thickness~$t$ with a dielectric constant~$\epsilon_1$ atop an infinitely thick dielectric layer of dielectric constant~$\epsilon_2$ has the form (see Appendix and Ref.~\cite{CNT})
\begin{equation}
U_{CP}(z)=-\ \frac{\hbar c\alpha_0}{2\pi}\ \frac{1}{z^4}\ F(\epsilon_1,\epsilon_2,t/z),
\end{equation}
where~$\alpha_0$ is the static atomic polarizability. The dimensionless function~$F$ takes the single-layer limiting value~$F\sim\frac{3}{4}~\frac{\epsilon-1}{\epsilon+1}~\phi(\epsilon)$ with~$\epsilon=\epsilon_1$ when~$z\ll t$, and with~$\epsilon=\epsilon_2$ when~$z\gg t$, where~$\phi(\epsilon)$ is on the order of unity~\cite{yan}. $F=\frac{3}{4}$~is obtained in the vicinity of a perfectly conducting thick layer. In our case~$\alpha_0=47.3\times10^{-24}\,\rm{cm^3}$ is the ground state static polarizability of the~$\rm^{87}Rb$ atom~\cite{antezza2}, $\epsilon_1=4$ for the~SiO$_2$ layer, and~$\epsilon_2=12$ for the~Si wafer.

\begin{figure}
 	\includegraphics[width=0.45\textwidth]{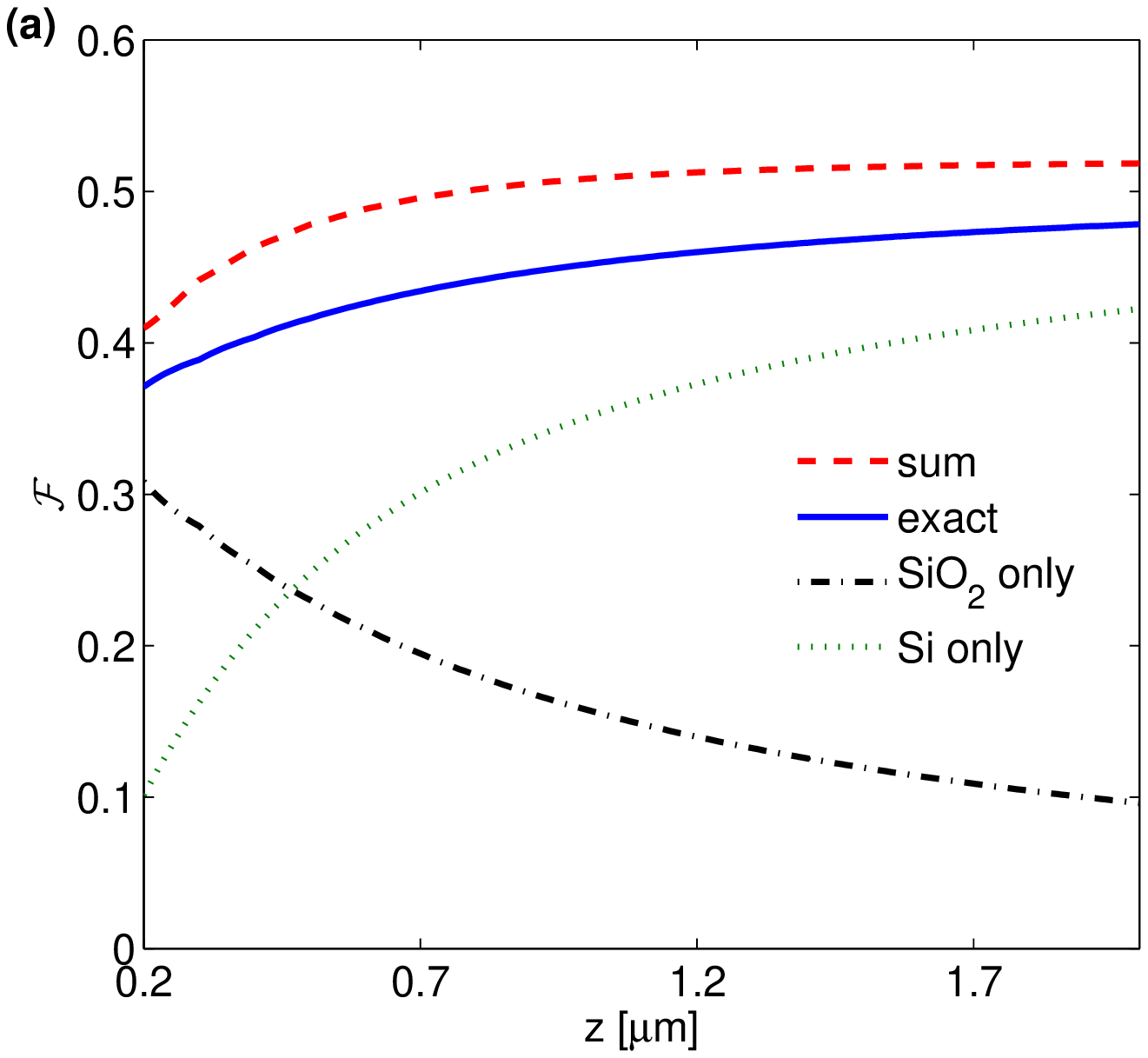}
	\includegraphics[width=0.45\textwidth]{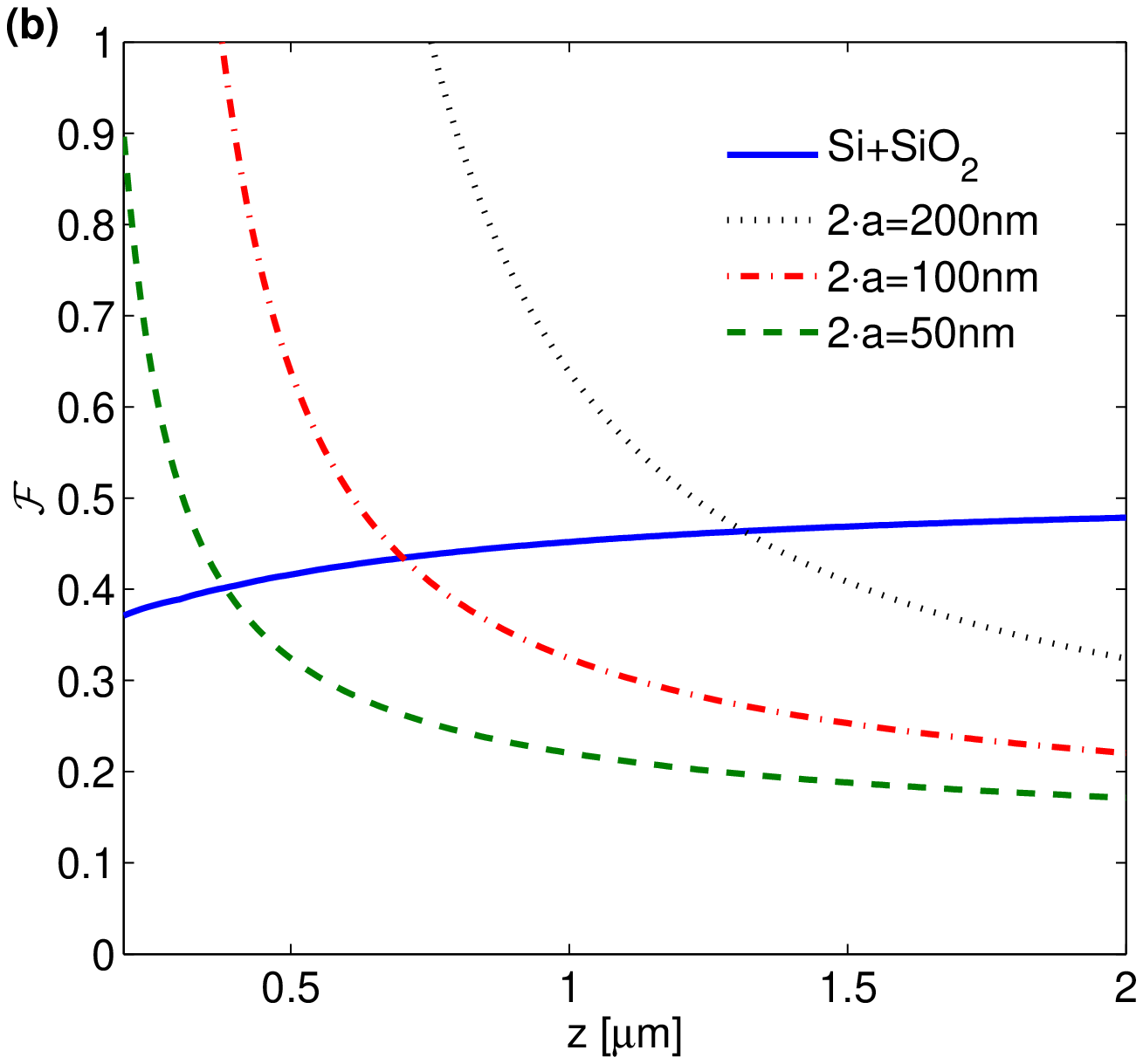}
\caption{(Color online) (a)~$\cal F$ factor for the bilayer system of thick~Si coated by a~$\rm100\,nm$ layer of~SiO$_2$, similar to the system studied in~Ref.~\cite{CNT}. The exact calculation (solid curve) is compared to the sum (dashed) of two separate systems~--~the~SiO$_2$ layer alone (dashed-dotted) and the~Si layer alone (dotted). For the contribution of the~Si layer the factor~$F$ would be constant for a system of coordinates starting at its top~($z=\rm-100\,nm$), but here it is rescaled to the coordinate system where~$z=0$ is at the top of the~SiO$_2$ layer (see text). The sum of the two separate contributions over-estimates the exact result by about~8-15\% over the relevant range. (b)~$\cal F$ factor (again rescaled to~$z=0$ at the top of the~SiO$_2$ layer) for the planar wafer (solid curve) reproduced from~(a) and for perfectly conducting cylindrical wires of diameters~$2a=\rm50-200\,nm$ (broken curves) lying on the wafer surface. Two important reasons for the differences between the wires are the different atom-wire distances~$z-2a$, which is smaller for thicker wires, and the larger solid angle subtended by the wider wire.}
\label{fig:cp}
\end{figure}

As stated above, we wish to compare contributions to the \CP\ potential from the three different components comprising the surface: the~Si chip, the~SiO$_2$ layer of thickness~$t$, and the gold nanowire of thickness~$h$. For this comparison to be meaningful, we require a common reference for the distance variable~$z$, which we define as the distance from the top of the~SiO$_2$ surface. Then the distance from the~Si chip is~$z+\rm100\,nm$ and the distance from the top of the gold nanowire is~$z-h$. To factor out the strong~$z^{-4}$ dependence, we plot the quantity~${\cal F}(z)\equiv-U_{CP}(z)~\frac{2\pi}{\hbar c\alpha_0}~z^4$ in Fig.~\ref{fig:cp}(a) for the~Si+SiO$_2$ bilayer. This is compared to a sum of two models (shown separately in the figure): one where the half space for~$z<\rm-100\,nm$ is full of~Si while the other half is empty; and another in which only a~$\rm100\,nm$-thick SiO$_2$ layer exists, with empty space for~$z<\rm-100\,nm$. The figure shows that simply summing the two potentials over-estimates the exact result by~8-15\% over the relevant range, but it gives the right order of magnitude.

 Next we consider the \CP\ potential for an atom at a distance~$R$ from the center of a cylindrical conducting wire of radius~$a$ where we set~$a=h/2$. It appears that the main contribution to the integral in Eq.~(\ref{cp_general}) comes from frequencies on the order of~$\omega\sim c/R$. In our case, where~$R\rm<1\,\mum$, the skin depth for a gold wire with resistivity~$\rho=\rm2.2\times10^{-8}\,\Omega\cdot m$ is~$\delta =\sqrt{2\rho/\mu_0\omega}\lesssim\rm10\,nm$, which is much smaller than the width or thickness of the nanowires considered. We can therefore use a model where the wire is perfectly conducting (impenetrable for~EM waves), such that the~EM Green's tensor is much simpler than in the general case. The \CP\ potential is then given by
\begin{equation}
U_{CP}(R)=-\ \frac{\hbar c\alpha_0}{2\pi}\ \frac{1}{(R-a)^4}\ F(a/R).
\end{equation}
For~$a/R>0.2$ the function~$F$ is nearly linear,~$F(a/R)\approx0.53(a/R)+0.22$, tending to~$F=\frac{3}{4}$ as~$R\rightarrow a$, where the surface of the cylinder is similar to a planar conducting surface. In the opposite limit~$a/R\ll0.1$ the function~$F$ drops to zero as~$F(a/R)\sim-\frac{2}{3\log(a/R)}$ (see Appendix).

Figure~\ref{fig:cp}(b) again shows the factor~$\cal F$ for the \CP\ potential from the planar~(\ie\ Si+SiO$_2$) surface in comparison to~$\cal F$ for cylindrical wires of different diameters~$2a$. It is evident that the contribution of the wire is dominant when the distance from the wire is less than~5-7 times the diameter of the wire. For larger distances the contribution of the wire falls to half or less than the contribution of the surface. Given our experience with the bilayer system~\cite{CNT}, we expect the exact calculation of the wire+surface to deviate by the same order as we observe for the bilayer, \ie\ less than~20\%. This degree of inaccuracy may also follow from the fact that the wires do not have circular cross-sections but square or rectangular ones. Therefore we believe that taking the sum of the two models can be expected to give at least an order of magnitude estimation of the \CP\ potential.

\section{Atom loss}\label{sec:loss}

In this section we analyze the lifetime for atoms in the nanowire trap. This lifetime includes the spin-flip rate due to thermally induced noise, the Majorana spin-flip rate, and the tunneling rate to the surface. Finally, we estimate the decoherence rate due to the thermally induced noise in the room temperature surface.

\subsection{Spin flip due to thermal noise\label{subsec:sf_sec}}

The magnetic thermal noise (Johnson noise) arising from conducting materials on the atomchip is coupled to the trapped atoms \via\ their magnetic moment~$\mu_A$. As a consequence, spin flips, heating and decoherence become dominant close to a conductor even without applied currents. Here we calculate the trap loss rate due to spin flips. We assume that the spectrum of magnetic noise from the conductor is roughly constant for frequencies in the~MHz region, the latter being able to drive magnetic transitions between Zeeman sub-levels in the same hyperfine level. In this case the magnetic moment is~$\bfmu_A=\mu_B g_F {\hat{\bf F}}$, where~$\mu_B$ is the Bohr magneton,~$g_F$ is the Land\'e factor for the hyperfine level and~${\hat{\bf F}}$ is the hyperfine spin operator. Using the theory developed by Henkel, the thermal spin-flip rate from an initial trapped Zeeman state~$|i\rangle$ to a final untrapped state~$|f\rangle$ can then be written as~\cite{Henkel,TalAni}:
\begin{equation}\label{eq:spinfliprate}
\Gamma_{\rm th}({\bf x}) =
\frac{\mu_B^2 g_F^2}{\hbar^2}
\sum_{j,k=\perp}\left\langle i\left|F_j\right|f\right\rangle
\left\langle f\left|F_k\right|i\right\rangle~
S_B^{jk}({\bf x},{\bf x},\omega_{if}),
\end{equation}
where we sum the contribution of all components of the noise perpendicular to the atomic magnetic moment. Here the function~$S_B^{jk}$ is the correlation function of the magnetic field noise, which is given by
\begin{equation}\label{eq:SB}
S_B^{jk}({\bf x}_1,{\bf x}_2,\omega\rightarrow 0)=
\frac{k_B T}{4\pi^2\rho\;\epsilon_0^2\;c^4}\
\left[{\rm Tr}\left\{X_{jk}({\bf x}_1,{\bf x}_2)\right\}\delta_{jk}-X_{jk}({\bf x}_1,{\bf x}_2)\right],
\end{equation}
with~$X_{jk}$ being a geometrical factor which also averages over~$1/\rho({\bf x})$ if the resistivity changes in space:
\begin{equation}\label{eq:Xij}
X_{jk}({\bf x}_1,{\bf x}_2) =
\frac{\rho}{2}
\int_V \frac{d^3{\bf x}'}{\rho({\bf x}')}\
\frac{\left(\mathbf{x}_1-\mathbf{x'}\right)_j
      \left(\mathbf{x}_2-\mathbf{x'}\right)_k}
     {\left|\mathbf{x}_1-\mathbf{x'}\right|^3
      \left|\mathbf{x}_2-\mathbf{x'}\right|^3}~.
\end{equation}
The~$X_{jk}$ sum up the contribution of local fluctuations arising from each point in the conductor's volume. We calculate Eq.~(\ref{eq:spinfliprate}) within the quasi-static approximation~\cite{TalAni,henkel2}, which is valid when the atom-conductor distance is smaller than the skin depth $\delta=\sqrt{2\rho/\mu_0\omega_L}$ ($\omega_L$ is the Larmor frequency). This condition is easily met here since the skin depths of metals in the~MHz region is typically tens of~$\mum$ (\eg\ gold has a skin depth of~$\delta\rm\approx70\,\mum$).

In Fig.~\ref{fig:lifetime}(a) we present estimated lifetimes for trapped atoms due to thermal noise-induced spin flips. The wire size greatly affects the lifetime, mostly because smaller wires place much less conducting material near the atoms, and also because of their higher resistivity. For a~$\rm50\times50\,nm$ wire, we estimate that the lifetime of a cloud trapped~$\rm500\,nm$ from the wire surface is~$\approx\rm5\,s$, so we do not expect thermal noise-induced spin flips to be a dominant loss mechanism in typical experiments.

\subsection{Majorana spin flips\label{subsec:majorana}}
Cold atoms in a low-field seeking state that are trapped near a vanishing magnetic field can undergo a spin-flip transition to a high-field seeking state that is untrapped (Majorana spin flips). Applying a small offset (Ioffe-Pritchard) field~$B_0$ will generate a non vanishing field at the trap center, hence reducing the spin-flip transition rate as given by the approximate formula~\cite{sukumar}:
\begin{equation}\label{eq:Majorana_spin_flip_rate}
    \Gamma_{\rm M}=\frac{\pi\omega_r}{2}~
    \exp\left(-\frac{2|\bfmu_A||\mathbf{B_0}|
    +\hbar\omega_r}{2\hbar\omega_r}\right),
\end{equation}
where~$\omega_r$ is the trap radial frequency. Equation~(\ref{eq:Majorana_spin_flip_rate}) is valid when the Larmor frequency $\omega_{\rm L}=|\bfmu_A||\mathbf{B_0}|/\hbar\gg\omega_r$, requiring that~$B_0\gg\rm50\,mG$ for typical radial frequencies. In the following sections, we choose a Ioffe-Pritchard field~$B_0$ that simultaneously satisfies this condition and yields a Majorana spin-flip lifetime of~$\rm2\,s$.

\subsection{Tunneling to the chip surface\label{subsec:tunneling}}

\begin{figure}[ht]
	\includegraphics[width=0.42\textwidth]{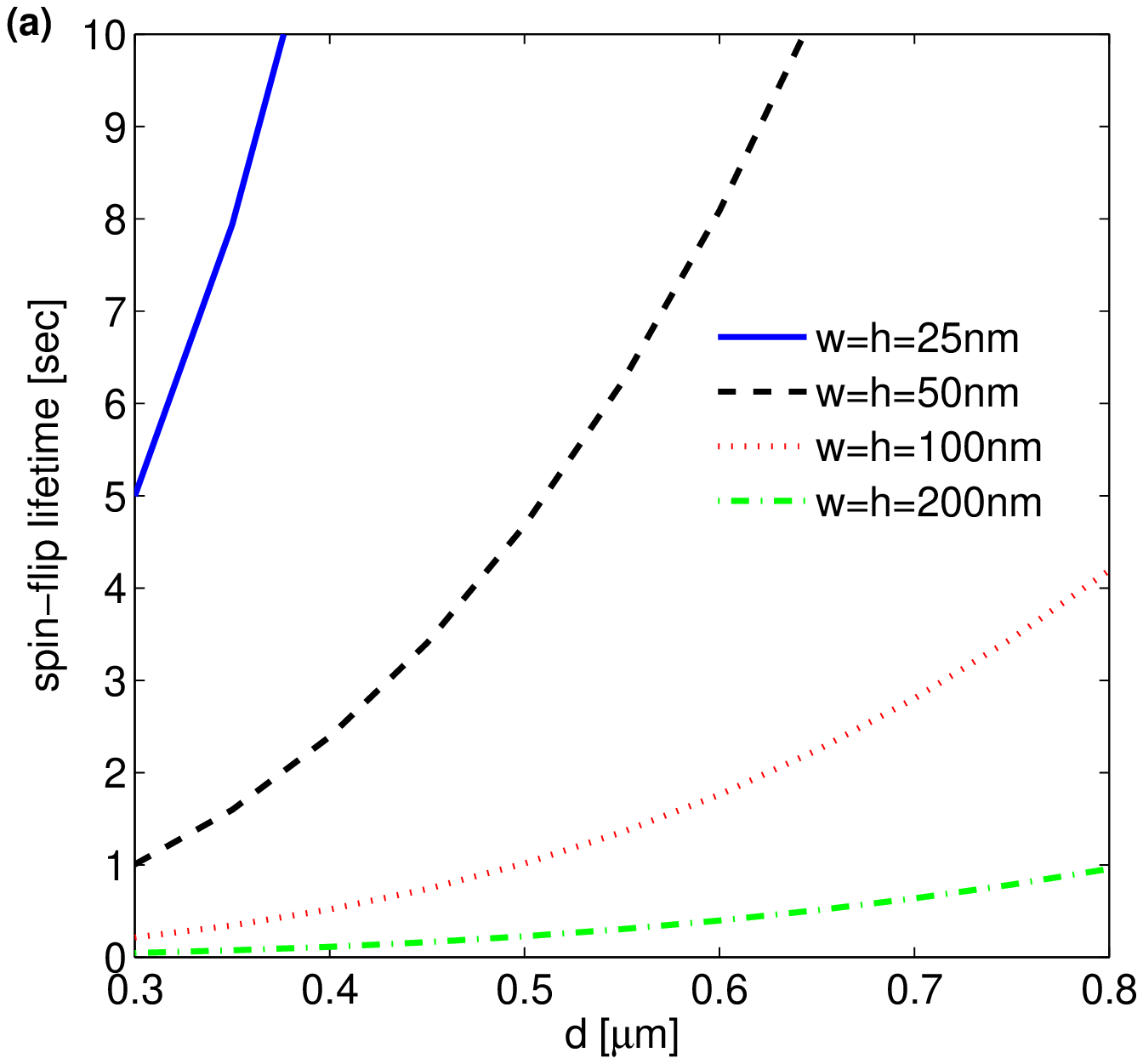}
	\includegraphics[width=0.42\textwidth]{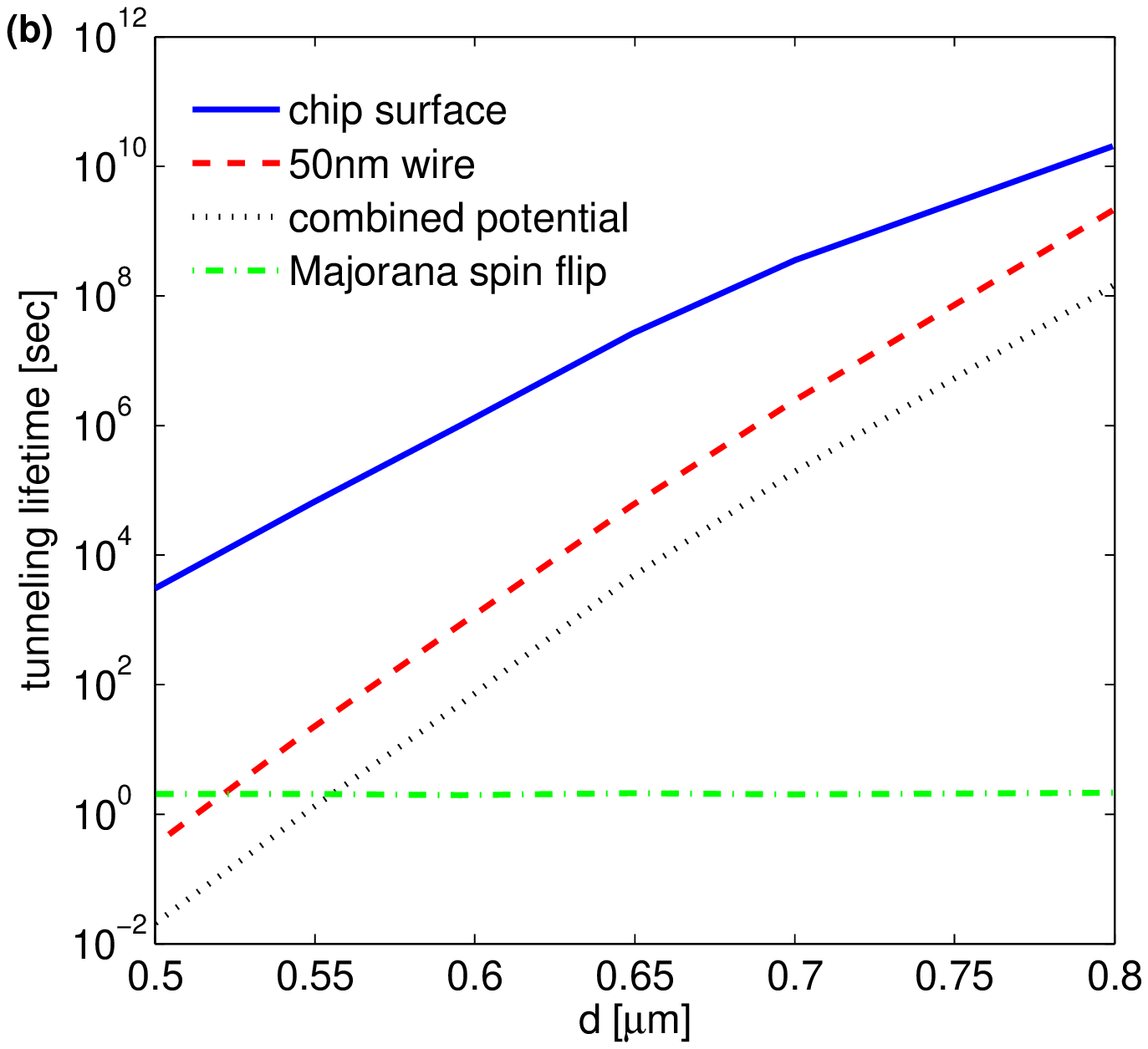}
	\caption{(Color online) (a)~Trap lifetimes due to thermal noise-induced spin flips, calculated for atoms trapped at distances~$d$ above wires with square cross sections of~$\rm25-200\,nm$. Reducing the wire size increases the lifetime; for the~$\rm50\times50\,nm$ wire the lifetime exceeds the selected Majorana lifetime of $\rm2\,s$ for distances~$d>\rm0.37\,\mum$. For comparison, the lifetime~$\rm1\,\mum$ above a very wide wire would be $<\rm10\,ms$. (b)~Tunneling lifetimes calculated for a~BEC of~1000 atoms in traps generated at different distances~$d$, compared to the Majorana spin-flip lifetime (kept constant at~$\rm2\,s)$, assuming a current of~$\rm40\,\muA$ passing through a~$\rm50\times50\,nm$ trapping wire. This current is more than an order of magnitude below the maximum for such a nanowire [Fig.~\ref{fig:rho}(c)]. The solid and dashed curves are calculated assuming surface-only and wire-only contributions to the \CP\ force respectively. Even though these \CP\ forces are of the same order of magnitude [Fig.~\ref{fig:cp}(b)], the atomic density is much higher directly above the wire, so tunneling to the wire is much faster than tunneling to the~Si+SiO$_2$ surface; the latter tunneling proceeds mostly from the cloud edges, where the atomic density is much lower. The dotted curve is calculated for a potential combining the wire and surface \CP\ forces; the corresponding tunneling lifetime is shorter yet because the trap barrier is reduced along the entire wire and at the cloud edges. In this approximate calculation, the tunneling lifetime exceeds the Majorana lifetime for distances~$d>\rm0.55\,\mum$. Using higher currents for such wires would increase the tunneling lifetime and is discussed further in Sec.~\ref{sec:example}.}
	\label{fig:lifetime}
\end{figure}

As a result of the \CP\ potential, the magnetic barrier between the surface and the atoms is lowered, and atoms can tunnel through the barrier to the atomchip surface or wire. Calculated tunneling lifetimes are presented in Fig.~\ref{fig:lifetime}(b), where we use a weighted average of the tunneling rate over all points in the~$(x,y)$ plane. For each point~$(x,y)$ we use the~WKB approximation for tunneling through a one-dimensional potential barrier along the~$z$ direction~\cite{CNT}:
\begin{equation}\label{eq:tunnel_rate}
    \Gamma_{{\rm tunn}}=\int\int dx\ dy\ P(x,y)\ \omega_r(x,y)
    ~\exp\left(-2\int_{z_{1}}^{z_{2}}dz
    \sqrt{\frac{2m}{\hbar^2}\left[U(x,y,z)-\mu\right]}
    \right),
\end{equation}
where~$\mu$ is the chemical potential and the integration over~$z$ is between the classical turning points~$z_1(x,y)$ and~$z_2(x,y)$ defined by $U(x,y,z_1)=U(x,y,z_2)=\mu$. The weighted tunneling probability appearing in the integrand is given at any point by~$P(x,y)=\frac{1}{N}\int dz\ n(x,y,z)$, where~$n(x,y,z)$ is the particle density and the transverse frequency $\omega_r(x,y)=\hbar \sqrt{\langle k_z^2\rangle}/2mL(x,y)$ is the inverse of the average round-trip time for a particle moving between the turning points [$L(x,y)=z_1(x,y)-z_2(x,y)$]. These quantities are all calculated by solving the \GP\ equation for~1000 atoms of~$\rm^{87}Rb$. In a typical trap generated by a~Z-shaped wire (\eg\ see Sec.~\ref{sec:example}), most of the tunneling occurs either at the center of the trap (where the atoms are closest to the wire) or at the trap ends (where the potential curves down towards the surface). Because of the much higher atom density directly above the wire, the lifetime is governed mostly by tunneling to the wire rather than to the surface, as discussed further in Sec.~\ref{sec:example}.

\subsection{Spatial decoherence\label{subsec:sp_decoherence}}

\begin{figure}[ht]
	\includegraphics[width=0.45\textwidth]{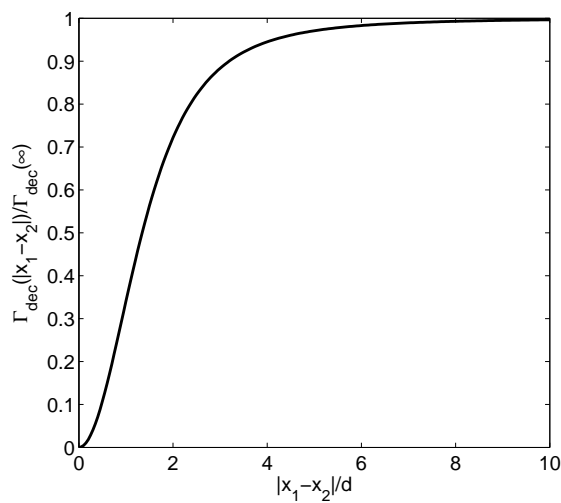}
	\caption{The rate of spatial decoherence between two points~${\bf x}_1$ and~${\bf x}_2$ above the wire as a function of their longitudinal separation, relative to their distance d to the wire. The decoherence rate is given relative to its maximum value when the two points are very far apart relative to $d$. The model assumes that the wire is infinitely long and much narrower than the distance~$d$. When the two points are separated by more than about~4 times their distance from the wire, the magnetic noise at the two points becomes uncorrelated, resulting in the decoherence rate asymptotically reaching a maximum as shown. Under these circumstances, the rate depends on the wire width in the same way as the thermal spin-flip lifetime shown in Fig.~\ref{fig:lifetime}(a), but is~2.4 times shorter (see text for details).}
	\label{fig:decoherence}
\end{figure}

As we have noted above, fluctuations of the magnetic field perpendicular to the quantization axis of the atom may cause transitions of the atom between Zeeman states defined along this axis. Conversely, fluctuations of the magnetic field along the quantization axis cause spatially dependent energy fluctuations, which may be viewed as potential fluctuations for the atom. These potential fluctuations imply that the phase of the atomic wavefunction at two locations~${\bf x}_1$ and~${\bf x}_2$ will also fluctuate, giving rise to dephasing. Again, we find that the mean square of the phase difference after a time~$t$ is given by
\begin{equation}
\langle [\phi({\bf x}_1)-\phi({\bf x}_2)]^2\rangle=
\frac{m_F^2g_F^2\mu_B^2}{\hbar^2}
\left\langle\left[\int_0^t dt' B_\parallel({\bf x}_1,t')
                 -\int_0^t dt''B_\parallel({\bf x}_2,t'')\right]^2\right\rangle,
\end{equation}
where~$B_\parallel$ is the magnetic field component along the quantization axis. For a time scale much longer than the inverse of the magnetic noise bandwidth, we may take the random-walk limit $\int_0^t dt' \int_0^t dt'' \langle B_\parallel(t')B_\parallel(t'')\rangle = \frac{1}{2}S^\parallel_B(\omega\rightarrow0)t$, where~$S^\parallel_B$ is the magnetic field correlation tensor of Eq.~(\ref{eq:SB}) along the coordinate of the quantization axis. It then follows that the square of the phase difference grows linearly with time. This implies, in accordance with the theory developed by Henkel~\cite{henkel1}, that the coherence, which may be defined as
$g^{(1)}({\bf x}_1,{\bf x}_2)=\langle e^{i[\phi({\bf x}_1)-\phi({\bf x}_2)]}\rangle$,
drops exponentially with the standard deviation of the difference, \ie\
$g^{(1)}({\bf x}_1,{\bf x}_2)=\exp(-\Gamma_{\rm dec}t)$, with
\begin{equation}
\Gamma_{\rm dec}=\frac{m_F^2g_F^2\mu_B^2}{2\hbar^2}
\left[ S^\parallel_B({\bf x}_1,{\bf x}_1)
     + S^\parallel_B({\bf x}_2,{\bf x}_2)
     -2S^\parallel_B({\bf x}_1,{\bf x}_2)\right]~.
\end{equation}
Figure~\ref{fig:decoherence} shows the decoherence rate of a split atomic wavefunction at two points located at an equal distance $d$ above an infinitely long and thin wire, as a function of the longitudinal separation between the points. Similar results are obtained when the two points are located above two separate parallel wires creating a local potential minimum above each of them, as a function of the separation between the two wires (and consequently between the two points). 
 When the distance between~${\bf x}_1$ and~${\bf x}_2$ is much larger than the distance to the wire, the correlation term~$S_B({\bf x}_1,{\bf x}_2)$ becomes negligibly small and the decoherence rate depends only on the distance of the two points from the wire. For $\rm^{87}Rb$ atoms in the state~$|F,m_F\rangle=|2,2\rangle$ it follows that for~${\bf x}_1, {\bf x}_2$ equidistant from the wire, the decoherence rate is $\Gamma_{\rm dec}(|{\bf x}_1-{\bf x}_2|\rightarrow \infty)=2.4\Gamma_{\rm th}$, where~$\Gamma_{\rm th}$ is given in Eq.~(\ref{eq:spinfliprate}). We see that, if the nanowire trap is constructed with a coherence lifetime long enough, \eg\ for interferometer experiments, then the experiment will not be limited by thermal spin-flip losses. Moreover, we see that coherence lifetimes on the order of~$\rm1\,s$ may be expected for the nanowire traps discussed here.

\section{Specific nanowire atomchip trap\label{sec:example}}

We now apply the foregoing general properties of nanowires and their associated magnetic fields and noise to a specific example.
We simulate a typical atomchip Z-shaped trap~\cite{RonRev}, aiming to achieve the smallest atom-surface separation while maintaining a lifetime~$>\rm1\,s$. Compatible with the fabrication process presented previously, we choose a~$\rm50\,\mum$-long gold nanowire with a~$\rm50\times50\,nm$ cross-section. This choice minimizes the \CP\ force due to the wire, thus lengthening the tunneling lifetime [Fig.~\ref{fig:cp}(b)], and at the same time reduces thermal magnetic noise contributions to atom loss [Fig.~\ref{fig:lifetime}(a)]. The ends of the nanowire are connected to conventional gold leads and are included in the magnetic field simulation. We consider a current of~$40\,\muA$, which is well below the calculated maximum current of~$\rm1.2\,mA$ [Fig.~\ref{fig:rho}(c)]. By applying a bias field of~$\rm132\,mG$ in the~$\hat{y}$ direction, a trap is generated at a distance of~$0.6\,\mum$ from the wire. Trap lifetimes for closer atom-surface distances would be limited by much faster tunneling. Applying a second bias field of~$\rm83\,mG$ in the~$\hat{x}$ direction ensures a Majorana spin-flip lifetime~$>\rm2.0\,s$. These parameters specify the basic atomchip trap configuration, whose properties we now discuss.

\begin{figure}[ht]
	\includegraphics[width=0.45\textwidth]{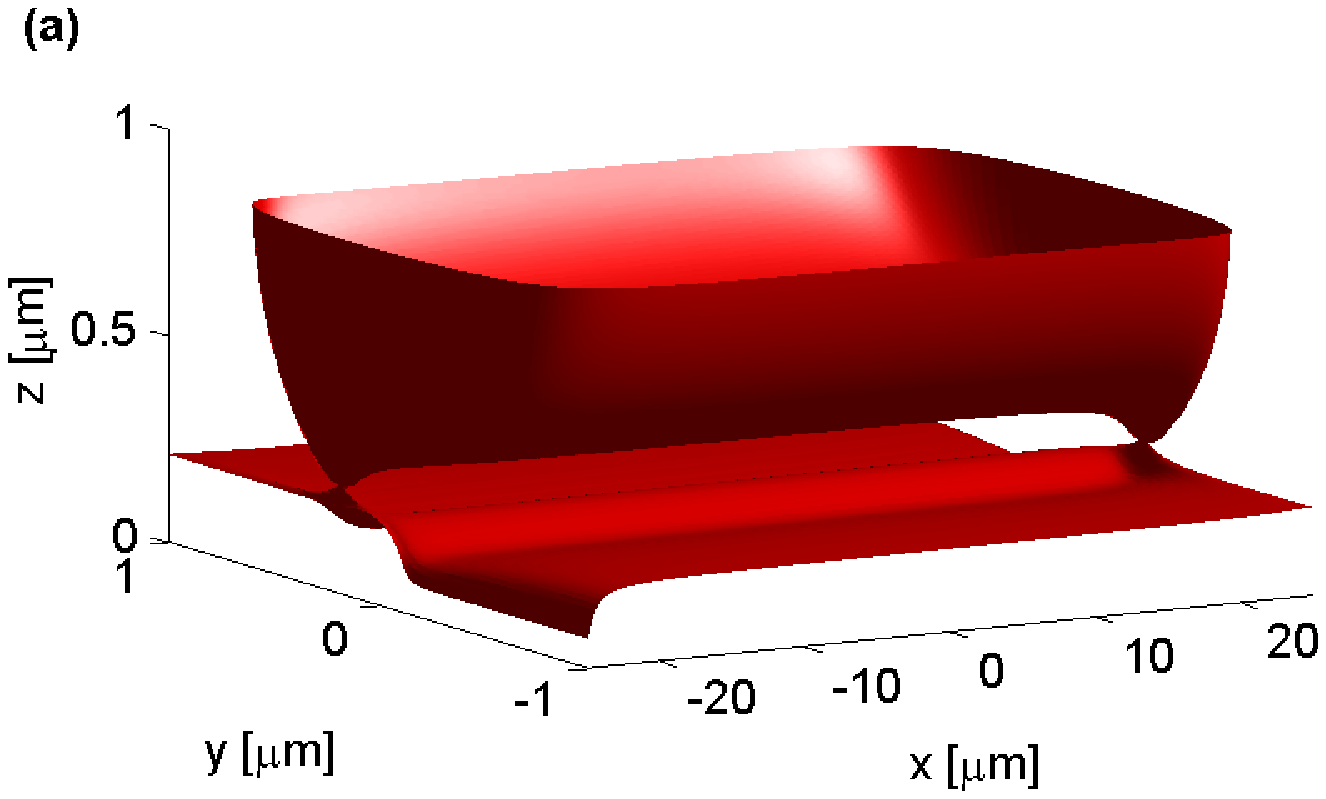}
	\includegraphics[width=0.45\textwidth]{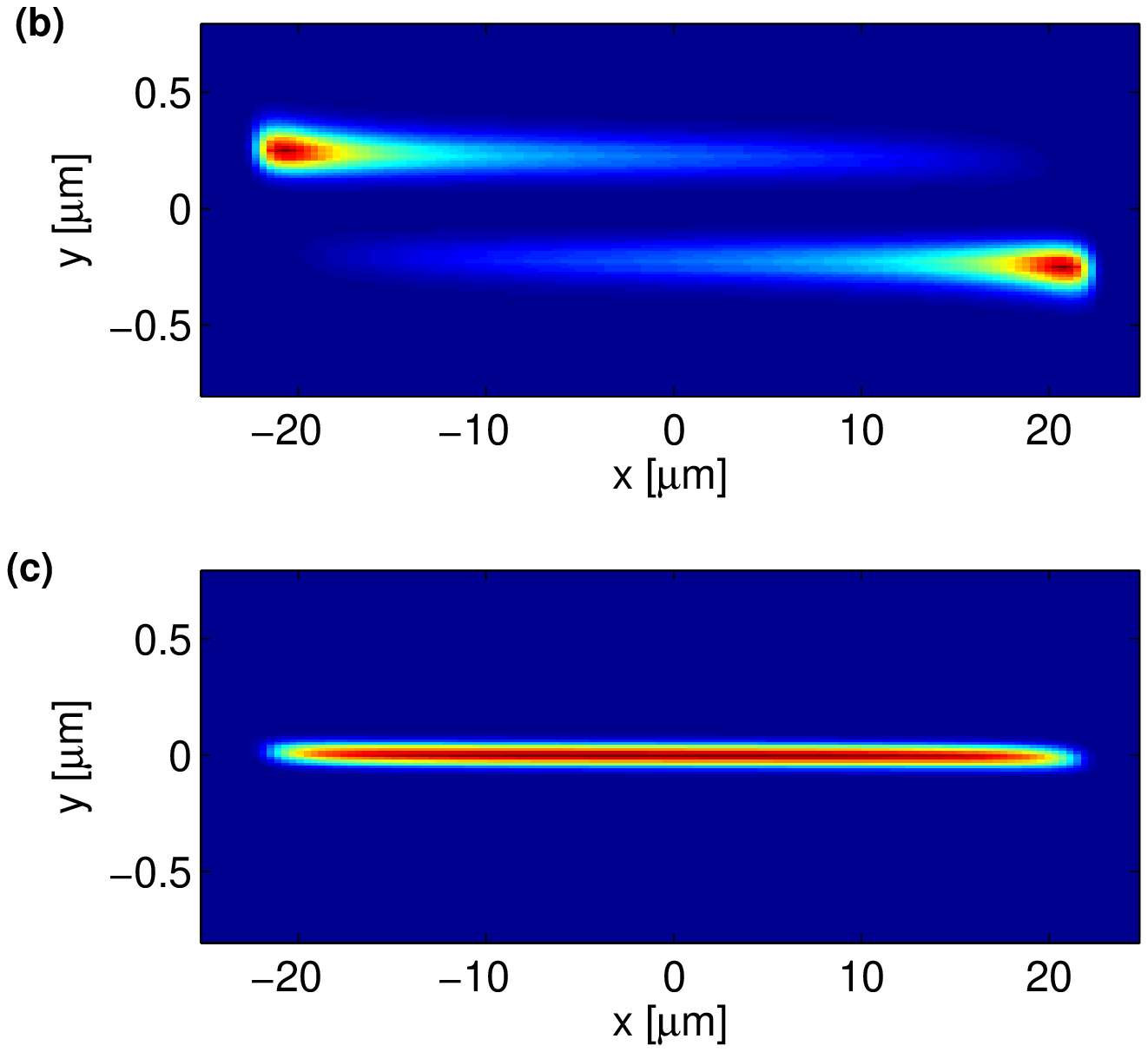}
	\caption{(Color online) (a)~Lower half of an isopotential surface at~$\rm2.9\,\muK$ for an atomchip trap centered at~$d=\rm0.6\,\mum$, created by passing a current of~$\rm40\,\muA$ through a~$\rm50\times50\,nm$ gold nanowire. The isopotential has been compressed in the $x$-direction by a factor of~20 for better visibility, and potential corrugations have been ignored. The nanowire is~$\rm50\,\mum$ long. The potential sheet below the closed trap surface is due to the \CP\ potential from the atomchip surface and the nanowire. The far edges of the isopotential surface touch the \CP\ potential sheet, implying that~$\rm2.9\,\mum$ is the trap depth. (b-c)~Weighted tunneling probabilities [integrand of Eq.~(\ref{eq:tunnel_rate})] calculated for a Bose-Einstein condensate of~1000 $\rm{^{87}}Rb$ atoms. The tunneling probabilities in~(b) consider only the \CP\ contribution from the Si+SiO$_2$ planar surface; in~(c) the calculation considers only the \CP\ contribution from the nanowire. Peak probabilities occur in~(b) at the ends of the trap because the potential bends towards the surface, even though the atomic density is low there. The peak tunneling probabilities in~(c) occur all along the wire axis, even though the potential barrier is higher there, since the atomic density is highest there also; the probabilities in~(c) are about ten times higher than in~(b). For the potential combining all \CP\ contributions, the tunneling lifetime is about~$\rm50\,s$.}
	\label{fig:isopotential}
	
\end{figure}

The trap depth, defined as the highest-energy isopotential that does not touch the surface, is about~$\rm2.9\,\muK$ and is limited by the \CP\ potential. An isopotential for a slightly higher energy is shown in Fig.~\ref{fig:isopotential}(a). The radial frequency at the trap minimum is about~$\rm10\,kHz$, which is controllable over a wide range since we are passing such a modest current through the nanowire. The connecting legs of the Z-wire act as ``end caps'' for a waveguide potential that lies almost directly above the nanowire. Corresponding weighted tunneling probabilities are shown in Fig.~\ref{fig:isopotential}(b) and~(c). As was shown in Fig.~\ref{fig:lifetime}(b), the tunneling lifetime due to the combined wire and surface \CP\ potential is an order of magnitude shorter than that due to the wire \CP\ potential. 

\begin{figure}[ht]
	\includegraphics[width=0.45\textwidth]{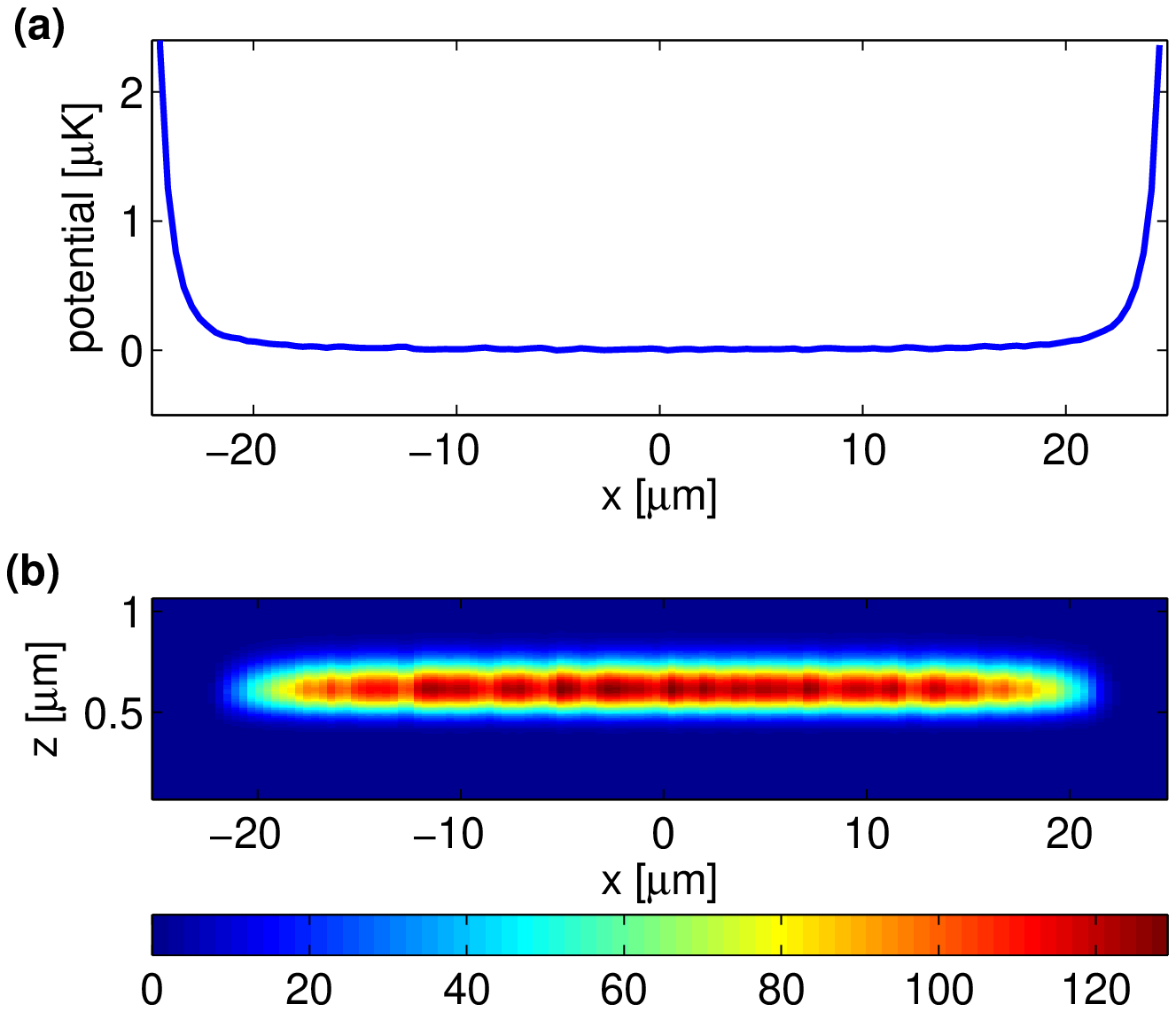}
	\includegraphics[width=0.45\textwidth]{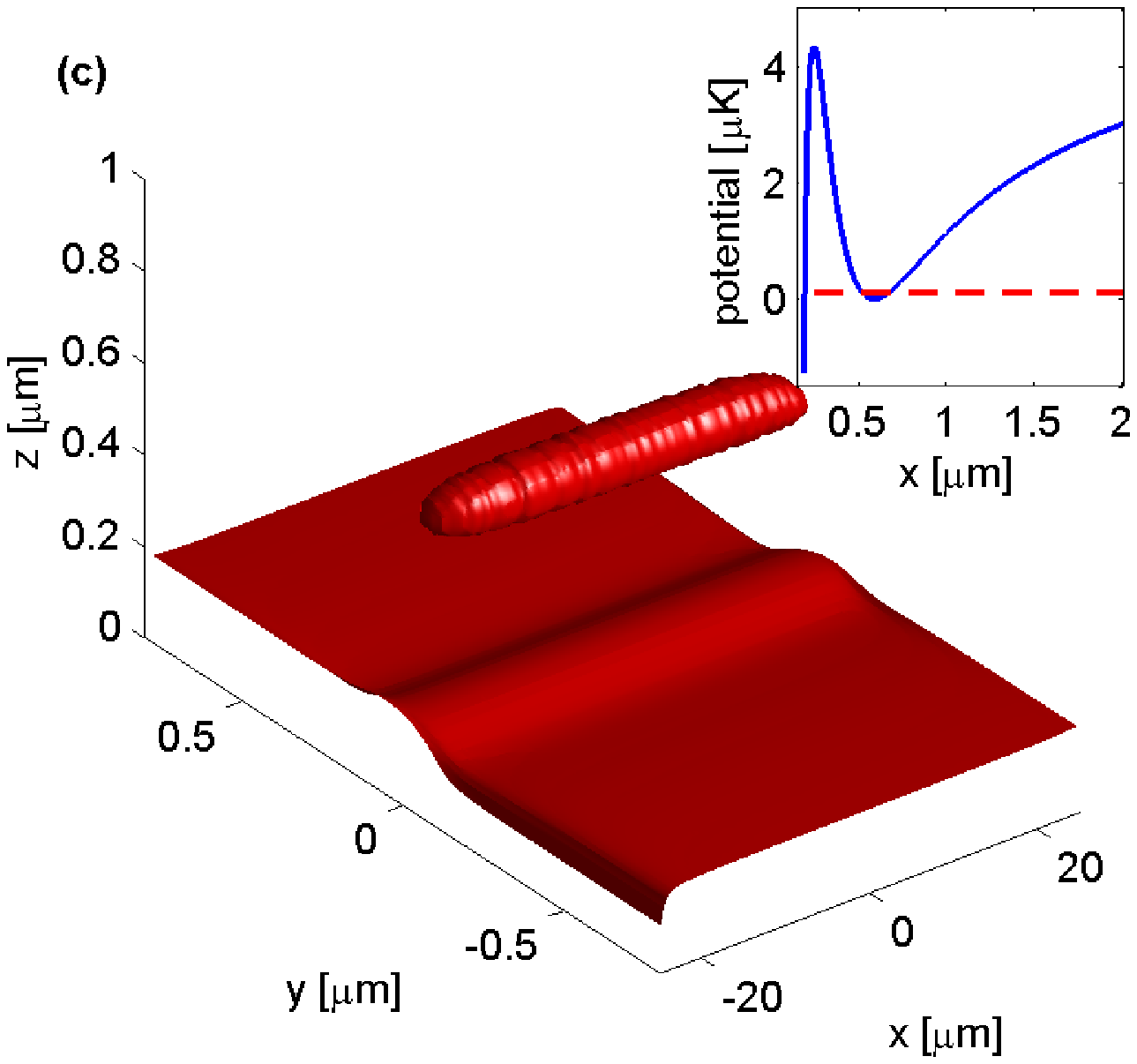}
	\caption{(Color online) (a)~Minimum-energy path for a trap generated by running~$\rm40\,\muA$ through a Z-shaped nanowire, where the central portion of the Z-wire is~$\rm50\,\mum$ long: for each value of the longitudinal distance~$x$, we show the minimum potential in the~$yz$-plane. (b)~Ground-state atomic density map (integrated along a viewing axis~$\hat{y}$ perpendicular to the nanowire and shown in units of $\rm\mum^{-2}$) for atoms at a trap height of~$d=\rm0.6\,\mum$, calculated with the \GP\ equation for~1000 atoms of~$\rm^{87}Rb$. The calculated potential corrugation perturbs the atomic density, which has a standard deviation of~3.8\% along its length. (c)~An isopotential surface of the trapping potential for an energy~$\rm0.1\,\muK$ above the trap minimum. The longitudinal axis is compressed by a factor of~50 for a better view. The potential corrugation is due to the trapping wire edge roughness, with a measured standard deviation of~$\rm2\,nm$. The potential sheet below the closed trap surface is due to the \CP\ potential from the atomchip surface and the nanowire. The inset shows a~1D cut of the potential above the trap center, with the energy of the isopotential surface shown by the dashed red line.}
	\label{Trap_Fig}
\end{figure}

The trap formed for such small distances from the nanowire is ``box'' shaped along the longitudinal axis, as shown in Fig.~\ref{Trap_Fig}(a). The smoothness of the trap bottom along the wire axis is limited only by the edge roughness of the nanowire (white-noise spectrum with~$\rm2\,nm$ rms, see Sec.~\ref{subsec:fab}) and has a standard deviation of about~${\rm8.2\,nK}$. The effect of this corrugation on the ground state of trapped atoms is examined by solving the \GP\ equation for~$N$ interacting bosons confined by the magnetic potential of the atomchip nanowire~\cite{dalfovo}:
\begin{equation}\label{eq:GP}
\left[-\frac{\hbar^2}{2m}\nabla^2+V(\mathbf{r})+g|\psi(\mathbf{r})|^2\right]\psi(\mathbf{r})=\mu\psi(\mathbf{r}),
\end{equation}
where~$m$ is the atomic mass,~$V(\mathbf{r})$ is the external potential,~$\mu$ the chemical potential, and~$g=4\pi\hbar^2a/m$ is the atom-atom coupling constant, with~$a$ being the {\it s}-wave scattering length ($a=\rm5.4\,nm$ for~$\rm^{87}Rb$). We do not use the Thomas-Fermi approximation since we do not expect a ``large'' number of atoms to be held in the trap. The calculated chemical potential for~$N=1000$ atoms of~$\rm^{87}Rb$ is~${\rm13\,kHz}\cdot2\pi\hbar$=~${\rm625\,nK}$, which is about~$\frac{1}{4}$ of the trap depth~ [Fig.~\ref{fig:isopotential}(a)]. In Fig.~\ref{Trap_Fig}(b) we present the calculated {\it in-situ} atomic density, which shows a standard deviation of~3.8\% due to the nanowire edge corrugation effects. The isopotential plotted in Fig.~\ref{Trap_Fig}(c) for an energy just above the minimum presents another view of the potential corrugation.

We see that a sufficiently deep trap can be formed using static magnetic potentials generated by nanowires, with sub-micron atom-surface separations. The main limitations for such traps will be cloud fragmentation and tunneling due to the \CP\ potential. The latter limitation can be overcome using higher currents. For a~$\rm50\times50\,nm$ wire, we can use a current up to~$\rm1.2\,mA$ [Fig.~(\ref{fig:rho})]; at a height~$d=\rm0.6\,\mum$ and a current~$I=\rm0.8\,mA$, the trap depth increases to~$>\rm100\,\muK$ and the tunneling lifetime increases by many orders of magnitude. However, in this configuration the potential corrugation causes severe atomic density fragmentation, with a calculated standard deviation of~40\%. In order to reduce the effects of potential corrugation, we can increase the chemical potential by shortening the nanowire trap. Shorter nanowires can actually be fabricated more easily and such traps would not be expected to significantly reduce the tunneling lifetime.

\section{Anisotropic conductors\label{sec:anisotropic}}
Replacing pure metals with other conductors such as superconductors, alloys or molecular conductors (see Introduction) may bring numerous advantages. As an example, we briefly describe utilizing electrically anisotropic materials and their qualities in the context of this work~\cite{TalAni}. In particular, if we orient the ``good'' conductivity~$a$-axis parallel to the wire direction~$\hat{x}$, the spatial and internal state decoherence rate of a trapped atomic sample is lowered by a factor on the order of the transverse conductivity suppression relative to gold, which may be several orders of magnitude. The greatest advantage of using such materials would therefore be for interferometric measurements with atomchips. The anisotropy in the resistance is also expected to decrease the effect of wire edge roughness on potential corrugations.

For a direct comparison with the gold Z-wire trap described above, we wish to maintain a current of~$\rm40\,\muA$ through the trap. The~$a$-axis resistivity for these materials is typically larger than the resistivity of gold. For a material such as~SrNbO$_{3.41}$ the temperature of the wire at a current density of~$\rm10^5\,A/cm^2$ is not expected to rise significantly~\cite{TalAni}, and hence the spin-flip lifetime due to thermal noise of the wire is expected to be about~200 times longer [Eq.~(\ref{eq:SB})] than for the comparable gold nanowire. We note that this improvement is not due to the anisotropic nature of the wire but simply to its high resistivity (about~200 times that of gold). An additional factor of~2 may be gained for ``one dimensional'' anisotropic conductors due to the specific anisotropic nature of these wires~\cite{TalAni}. Reducing the current density to~$\rm10^5\,A/cm^2$ (2 orders of magnitude smaller than that allowed in gold) requires a correspondingly increased wire cross-section of~$\rm200\times200\,nm$ to maintain the same current. Following Fig.~\ref{fig:lifetime}(a) this would reduce the above lifetime by a factor of about~20, ultimately yielding an increased spin-flip lifetime of a factor of~$\sim20$.

Metallic nanowires have limited maximum current densities due to the increase in their wire resistivity (Fig.~\ref{fig:rho}) from diffusive scattering at the wire surfaces. However, surface scattering for an anisotropic wire (where the surfaces are parallel to the good conductivity axis) may be significantly smaller and will have less effect on the wire resistivity, hence enabling higher current densities than discussed above. One may even speculate that at small dimensions the resistance relevant for the current density [Eq.~(\ref{eq:sondheimer})] and that relevant for the Johnson noise [Eq.~(\ref{eq:SB})] may become decoupled.

It is evident that the behavior of electrons in anisotropic materials in the context of surface scattering (resistance) and edge currents (fragmentation) requires further theoretical and experimental study. In any case, it appears that utilizing anisotropic materials could further improve the advantages of nanowires. Specifically, it may improve the coherence time by several orders of magnitude, and could reduce potential corrugations~\cite{TalAni}. Fabrication protocols for such anisotropic nanowires are being pursued in our laboratory.

\section{Summary and Conclusions}\label{sec:conclusion}

We have presented an analysis for further miniaturization achievable in atomchips based on current carrying wires, aiming to create static magnetic potentials capable of manipulating atoms on the scale of their deBroglie wavelength. We have analyzed the physical limitations of conducting wires, and we have also analyzed limiting effects due to the nearby surface, explicitly considering tunneling to the surface, surface thermal noise (causing both spin flips and decoherence), electron scattering within nanowires causing static potential corrugations, and the \CP\ force. Additional effects such as Majorana spin flips have also been taken into account.

We have analyzed a specific example of a nanowire trap, utilizing a standard configuration. We have shown that when utilizing nanowires, the main limitations to trapping atoms at sub-micron atom-surface distances are potential corrugation and tunneling to the surface. We briefly described an anisotropic conductor as a potentially useful alternative to standard gold wires. These examples serve not only to summarize the more general statements of the paper, but also as an outlook for further work which may include alternative geometries and materials.

We have shown that further miniaturization of atomchips, utilizing wires with cross sections as small as a few tens of nanometers, enables robust operating conditions for atom optics. Such miniaturization may allow the realization of potentials (\eg\ tunneling barriers) with a scale of the deBroglie wavelength, thereby bringing the atomchip a step closer to fulfilling its promise of a compact device for complex and accurate quantum optics with ultracold atoms. Achieving such small atom-surface distances should also contribute to the study of fundamental surface phenomena.

\section{Appendix: Casimir-Polder potential -- derivation}

\newcommand{\bea}{\begin{eqnarray}}
\newcommand{\eea}{\end{eqnarray}}
\newcommand{\beq}{\begin{equation}}
\newcommand{\eeq}{\end{equation}}
\subsection{A planar bilayer surface}

We consider a planar structure with a dielectric function given by
\beq \epsilon(z)=\left\{\begin{array}{lc} 1, & z>0, \\ \epsilon_1, & -b<z<0, \\
\epsilon_2, & z<-b. \end{array}\right.
\eeq
The Green's tensor for the EM field may be
derived from the reduced Green's tensor, which can be written as a sum
over transverse modes with well defined transverse wave vectors
${\bf k}=(k_x,k_y)$
\beq \bm{\Gamma}({\bf r},{\bf r}',t-t')=\int_{-\infty}^{\infty}
\frac{d\omega}{2\pi}
e^{-i\omega(t-t')}\int \frac{d^2{\bf k}}{(2\pi)^2} e^{i[k_x(x-x')+k_y(y-y')]}
{\bf g}_{\bf k}(z,z'),
\eeq
where ${\bf g}_{\bf k}(z,z')$ in the region $z,z'>0$ can be written
in terms of transverse electric (TE) and transverse magnetic (TM) scalar
Green's functions
\beq g^{\rm TE,TM}_{\bf k}(z,z')=\frac{i}{2k_z}\left[e^{ik_z|z-z'|}
+R^{\rm TE,TM}e^{ik_z|z+z'|}\right],\label{tetmg}
\eeq
with $k_z^2=\omega^2/c^2-k^2$. The reflection coefficients $R^{\rm TE,TM}$
are given below.
The first term in Eq.~(\ref{tetmg}) is irrelevant, being cancelled by the
vacuum subtraction in Eq.~(\ref{cp_general}).
The trace over the remaining part of the Green's tensor is now given in terms
of the (vacuum subtracted) scalar
functions $g^{\rm TM}$ and $g^{\rm TE}$ at $z=z'$ as
\beq \sum_i g^{ii}_{\kappa}=\omega^2 g^{\rm TE}+(k^2-k_z^2)g^{\rm TM}. \eeq

We now make the transformation $\omega/c \rightarrow i\zeta$ such that
$k_z\rightarrow i\kappa$
with $\kappa^2=k^2+\zeta^2$. The reflection coefficients $R^{\rm TE}$ and
$R^{\rm RM}$ are now given by
\beq
R^m=\frac{r^m_1+r^m_{12}e^{-2\kappa_1 b}}{1+r^m_1 r^m_{12}e^{-2\kappa_1 b}},
\label{Rm}
\end{equation}
for $m={\rm TE,TM}$, where
\begin{equation} r^{\rm TE}_{12}
=\frac{\kappa_1-\kappa_2}{\kappa_1+\kappa_2},\quad
r^{\rm TE}_{1}=\frac{\kappa-\kappa_1}{\kappa+\kappa_1},
\label{rTE}
\end{equation}
and
\begin{equation} r^{\rm TM}_{12}=\frac{\kappa_1/\epsilon_1-\kappa_2/\epsilon_2}
{\kappa_1/\epsilon_1+\kappa_2/\epsilon_2},\quad
r^{\rm TM}_{1}=\frac{\kappa-\kappa_1/\epsilon_1}{
\kappa+\kappa_1/\epsilon_1},
\label{rTM}
\end{equation}
with $\kappa_i=\sqrt{\epsilon_i\zeta^2+k^2}$.
We now obtain for the CP potential
\beq U_{CP}(z)=-\hbar c\int_{-\infty}^\infty
d\zeta\ \alpha(i\zeta)\int \frac{d^2{\bf k}}{(2\pi)^2}
\frac{1}{2\kappa}\left[(2k^2+\zeta^2)R^{\rm TM}-\zeta^2R^{\rm TE}\right]
e^{-2\kappa z}.\label{cppot}
\eeq
The formulas for the multilayer Green's functions have appeared in
many places, for example in Ref.~\cite{klim}, and the result~(\ref{cppot})
was first derived by Zhou and Spruch~\cite{zhou}.

At a distance $z$ which is much larger than $c/\omega_0$, where $\omega_0$ is
the lowest
optical transition frequency of the atom, we may assume that
$\alpha(\omega)\sim \alpha(0)$.
We now make the transformation $\kappa=p/z$ and $\zeta=\kappa\mu$ and obtain
\beq U_{CP}(z)=-\frac{\hbar c\alpha(0)}{2\pi z^4}F(\epsilon_1,\epsilon_2,b/z),
 \eeq
with
\beq F=\int_0^{\infty}dp\ p^3 e^{-2p}\int_{-1}^{1}d\mu
\left[\left(1-\frac{\mu^2}{2}\right)R^{\rm TM}-\frac{\mu^2}{2} R^{\rm TE}
\right].
\eeq
where $R^{\rm TM}$ and $R^{\rm TE}$ are given in Eqs.~(\ref{Rm})--(\ref{rTM})
with
$\kappa_i$ replaced by $p_i=p\sqrt{1+\mu^2(\epsilon_i-1)}$ and $b$ in the exponent
replaced by $b/z$.

When $b\ll z$ the exponent $\exp[-2p_1 b/z]\rightarrow 1$ and one obtains
$R^{\rm TE}=(p-p_2)/(p+p_2)$ and $R^{\rm TM}=(p-p_2/\epsilon_2)
(p+p_2/\epsilon_2)$, which are the reflection coefficients
for an interface between vaccum and a medium with dielectric function
$\epsilon_2$. On
the other hand, when $b \gg z$ we obtain $R^{\rm TE}\rightarrow r_1^{\rm TE}$
and $R^{\rm TM}\rightarrow r_1^{\rm TM}$. We conclude that the CP potential
becomes similar to that
generated by the deeper layer when $z\gg b$ and similar to the one generated by
the outer layer when $z\ll b$.

\subsection{Cylindrical wire}

The Green's tensor in cylindrical coordinates is given by \cite{milton}
\bea \bm{\Gamma}({\bf r},{\bf r}',\omega)=\sum_{m=-\infty}^{\infty}
\int_{-\infty}^\infty
{\frac{dk}{2\pi}}
\left[-\hat{{\bf M}}\hat{{\bf M}}'^*
\frac{(\hat{d}_m-k^2)}{\omega^2}F_m(r,r') \right. \nonumber \\
\left. \mbox{}+\hat{{\bf N}}\hat{{\bf N}}'^*\frac{1}{\omega}G_m(r,r')\right]
\chi_{mk}(\phi,z)\chi^*_{mk}(\phi',z'), \eea
where $\hat{{\bf M}}$ and $\hat{{\bf N}}$ are vectorial differential
operators that generate
the vector fields from the scalar fields for the TE and TM modes, respectively,
$\hat{d}_m$ is the two-dimensional Laplacian operator where the
differentiation with respect to $\phi$ is replaced by
$\partial/\partial \phi\rightarrow im$,
$F_m(r,r')$ and $G_m(r,r')$ are the scalar radial Green's function for the
TE and TM
modes, respectively, and $\chi_{mk}=(2\pi)^{-1/2}e^{ikz}e^{im\phi}$ is the
wave function
that holds the angular and longitudinal dependence of each mode. As we are interested
only in the single-point case $\phi=\phi'$ and $z=z'$, we have here
$\chi_{mk}\chi_{mk}^*=1/2\pi$.

\begin{figure}[ht]
	\includegraphics[width=0.45\textwidth]{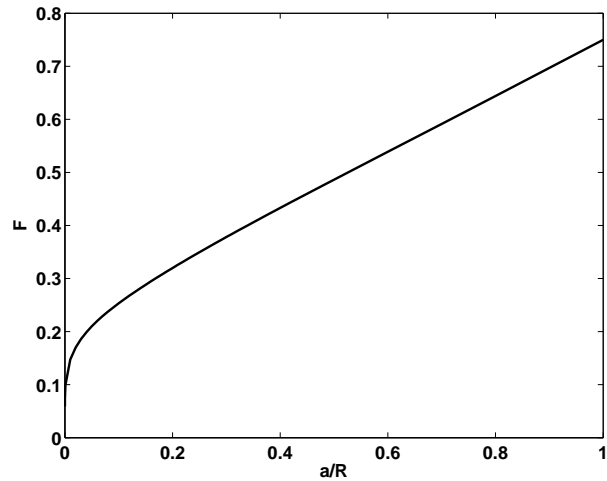}
	\caption{The function $F(a/R)$ for a perfectly conducting cylindrical wire.}
	\label{fig:cylF}
\end{figure}

By substituting for the right forms of the Green's function we obtain
\bea
\sum_j \Gamma_{jj}(R,R)&=&\frac{1}{2\pi}\sum_{m=-\infty}^{\infty}
\int_{-\infty}^{\infty}{
\frac{dk}{2\pi}}\frac{i\pi}{2\lambda^2} \left[-\omega^2\frac{m^2}{R^2}
\frac{J'_m(\lambda a)}{H'_m(\lambda a)}H_m^2(\lambda R)
+\omega^2\lambda^2\frac{J'_m(\lambda a)}{H'_m(\lambda a)}{H'_m}^2(\lambda R)
\right. \nonumber \\
&&\quad\left.
\mbox{} -\left(\frac{m^2 k^2}{R^2}+\lambda^4\right)\frac{J_m(\lambda a)}
{H_m(\lambda a)}
H_m^2(\lambda R)-k^2\lambda^2\frac{J_m(\lambda a)}{H_m(\lambda a)}
{H'_m}^2(\lambda R)\right],
\eea
where $J_m$ and $H_m$ are the Bessel functions and the Hankel functions of
the first kind
and $\lambda^2=\omega^2/c^2-k^2$ is the wave-vector along the radial direction.
As above, we
now make the transformation $\omega/c\rightarrow i\zeta$, such that
$\lambda\rightarrow
i\kappa$ with $\kappa^2=k^2+\zeta^2$. The Bessel function $J_m(\lambda a)$
transforms
into $I_m(\kappa a)$, where $I_m$ is the modified Bessel function, and
$H_m$ transforms
into $K_m$, the modified Bessel function of the second kind. We then obtain the
following result:
\bea
U_{CP}(R)&=&-\frac{\hbar c}{2\pi}\sum_{m=-\infty}^{\infty}
\int_{-\infty}^\infty{d\zeta \alpha(i\zeta)
\int_{-\infty}^\infty{\frac{dk}{2\pi}}}
\frac{1}{\kappa^2}\left\{-\frac{I'_m(\kappa a)}{K'_m(\kappa a)}\left(\frac{\zeta^2 m^2}{R^2}
K_m^2(\kappa R)+\kappa^2\zeta^2K'_m(\kappa R)^2\right)\right. \nonumber \\
&&\quad\left. +\frac{I_m(\kappa a)}{K_m(\kappa a)}\left[\left(\kappa^4+\frac{m^2 k^2}{R^2}
\right)K_m^2(\kappa R)+\kappa^2k^2 K'^2_m(\kappa R)\right]\right\}.
\eea
Now we take the static approximation for the polarizability $\alpha(i\zeta)\rightarrow
\alpha(0)$ and change variables to $x=\kappa R$, $\zeta=(x/R)\cos\phi$ and
$k=(x/R)\sin\phi$ and find
\bea
U_{CP}(R)&=&-\frac{\hbar c\alpha(0)}{2\pi R^4}\sum_{m=-\infty}^{\infty}
\int_0^\infty dx
\left\{-\frac{I'_m(xa/R)}{K'_m(x a/R)}\left(\frac{x m^2}{2}
K_m^2(x)+\frac{x^3}{2} K'_m(x)^2\right)\right. \nonumber \\
&&\left. +\frac{I_m(x a/R)}{K_m(x a/R)}\left[\left(x^3
+\frac{m^2 x}{2}
\right)K_m^2(x)+\frac{x^3}{2} K'^2_m(x)\right]\right\}.
\eea
This can be written in the form
\beq U_{CP}(R)=-\frac{\hbar c\alpha(0)}{2\pi (R-a)^4}F(a/R),
\eeq
with
\bea
F(a/R) &=& \left(1-\frac{a}{R}\right)^4\sum_{m=-\infty}^{\infty}\int dx\ x\
\left[\frac{I_m(x a/R)}{K_m(x a/R)}x^2K_m^2(x) \right. \nonumber \\
&&\left.\quad \mbox{}+\frac{1}{2}\left(\frac{I_m(x a/R)}{K_m(x a/R)}
-\frac{I'_m(xa/R)}{K'_m(x a/R)}\right)
\left(m^2 K_m^2(x)+x^2 K'_m(x)^2\right)\right].
\eea
The result of a numerical integration of $F$ is shown in Fig.~\ref{fig:cylF}.
It is found that
at $a/R\rightarrow 1$, such that the atom is very close to the surface relative
to the radius
$a$, we obtain the same result as for a plane conductor $F=3/4$. In the other
limit, where
$a/R\rightarrow 0$ one may see that $F\sim-\frac{2}{3\log(a/R)}$ and the
contribution to $F$ is dominated by the term $m=0$ only.

\begin{acknowledgments}
We thank the team of the Ben-Gurion University Weiss Family Laboratory for Nanoscale Systems for the fabrication and characterization of the nanowires and the BGU AtomChip group for fruitful discussions. It is a pleasure to thank Carsten Henkel (Potsdam) for discussion and critical comments. This work was supported by the European Community ``Atomchip'' Research Training Network and the EC Marie-Curie programme, the American-Israeli Binational Science Foundation (BSF), the Israeli Science Foundation, the French Embassy in Israel, the German-Israeli Binational Science Foundation (GIF), and the Ministry of Immigrant Absorption (Israel). The work of K.A.M. is supported in part by grants from the US Department of 
Energy and the US National Science Foundation.
\end{acknowledgments}


\end{document}